\documentstyle[preprint,aps,eqsecnum,epsfig]{revtex}
\begin{document}
\preprint{CMU-HEP-97-07; DOE-ER/40682-132} 
\draft 
\title{\bf NON-EQUILIBRIUM PRODUCTION OF PHOTONS VIA $\pi^0\rightarrow2\gamma$
IN DCC's} 
\maketitle
\begin{center}
{\it{\bf D. Boyanovsky$^{(a)}$, H. J. de Vega$^{(b)}$,
R. Holman$^{(c)}$, S. Prem Kumar$^{(c)}$}}
\end{center}
\begin{center}
{\it{(a) Department of Physics and Astronomy, University of Pittsburgh,
PA.\\15260,U.S.A.}}
\end{center}
\begin{center}
{\it{(b) LPTHE, Universite' Pierre et Marie Curie (Paris VI), Tour 16,
1er. etage,\\4, Place Jussieu 75252 Paris, Cedex 05, France}}\\
\end{center}
\begin{center}  
{\it{(c) Department of Physics, Carnegie Mellon
University, Pittsburgh, PA.15213, U.S.A.}}
\end{center}
\begin{abstract}
We study production of photons via the non-equilibrium relaxation of a
Disoriented Chiral Condensate with the chiral order parameter having a large
initial amplitude along the $\pi^0$ direction. Assuming the validity of the low
energy coupling of the neutral pion to photons via the $U_A(1)$ anomalous
vertex, we find that for large initial amplitudes along the $\pi^0$ direction,
photon production is enhanced by parametric amplification. These processes are
non-perturbative with a large contribution during the non-equilibrium stages of
the evolution and result in a distinct distribution of the produced photons and
a polarization asymmetry.  For initial amplitudes of the $\pi^0$ component of
the order parameter between 200-400 MeV, corresponding to energy densities
between 1-12 $\text{GeV}/\text{fm}^3$ we find a peak in the photon distribution
at energies between $\approx 300 -600$ MeV. We also find polarization
asymmetries typically between $5-10 \%$. We discuss the potential experimental
impact of these results as well as the problems associated with its detection.

\end{abstract}
\section{introduction}

The possibility that a disoriented chiral condensate (DCC) might form during
the early stages of a high energy, heavy-ion collision, has attracted
considerable attention in recent years \cite{anselm1}. The essential idea is
that, near the central rapidity region, large energy densities $\sim$ a few
GeV/$\text{fm}^3$ are achieved, corresponding to temperatures above 200 MeV, at
which the chiral symmetry is restored\cite{bjor}.  As this region cools down
via hydrodynamical expansion, it is conceivable that the QCD order parameter
might be trapped in a configuration where it points in a direction different
from the $\sigma$ direction which corresponds to the true vacuum. Subsequent
relaxation of such a DCC configuration to the true vacuum is expected to be
accompanied by a strong coherent burst of pions, which would then provide a
definitive experimental signature of the chiral phase transition.  The
formation and relaxation of these DCC configurations could potentially provide
an explanation for the CENTAURO and JACEE events found in cosmic ray emulsion
experiments\cite{bjor}.

Recent theoretical investigations \cite{wilraj,randrup,cooper,boyan} of
effective field theories of QCD, particularly the linear sigma model seem to
indicate that there is the possibility that such coherent regions may be formed
during the non-equilibrium stages of the chiral phase transition. Results of
classical\cite{wilraj}, semiclassical\cite{randrup} and quantum calculations,
with\cite{cooper} and without\cite{boyan} expansion seem to indicate that there
is a range of initial conditions for which the chiral phase transition goes
through a non-equilibrium stage during which coherent pion clouds like DCC's
may be formed.

The telltale experimental signal for  DCC events would be a non-binomial
distribution for the ratio of the number of neutral pions to the total
number of pions $R= n_{\pi^0}/(n_{\pi^0}+n_{\pi^+ \pi^-})$ with a
probability distribution $P(R)= 1/(2\sqrt{R})$\cite{bjor}. Both MINIMAX
at the Fermilab Tevatron and WA98 at CERN are currently looking for 
strong fluctuations in the ratio of neutral to charged pions, and
STAR and PHENIX at RHIC are scheduled to search for such events.    

There are potential ambiguities in obtaining clear signals for the probability
distribution $P(R)$.  Since the emitted pions are strongly interacting objects
they will undergo strong final state interactions, and it could be that a DCC
signal may be distorted and become indistinguishable from the background. An
important limitation for the determination of $P(R)$ is the accounting of
neutral pions which decay to two photons before reaching the detector.  The
neutral pion distribution is typically reconstructed from the analysis of the
emitted photons, a very demanding analysis in typical heavy-ion collisions with
large multiplicity events.
 
It is thus important to identify alternative signatures of the chiral phase
transition, and DCC formation and relaxation. Electromagnetic probes such as
photons and dileptons are ideal candidates since their interactions are weak,
and hence they can carry the information from the early stages of the evolution
of the plasma with minimal distortion.  It is very likely that to extract
exotic signals for the chiral phase transition or DCC events, both hadronic and
electromagnetic probes will be needed.

Furthermore, with the possibility of non-equilibrium effects during the
formation and expansion of the fireball in heavy ion collisions, the
study of unusual non-equilibrium processes may provide a greater understanding
of 
the signatures  of the Quark-Gluon and Chiral
phase transitions in experiments at RHIC and LHC. 

In an earlier work \cite{photop}, we have recently studied direct photon
production enhancement from the non-equilibrium relaxation of DCC's. There, we
studied the gauged linear sigma model and concentrated on the {\em
electromagnetic interactions of the hadronic current}. These calculations
involved the electromagnetic interactions of the charged pions alone. The
conclusion of this study was that the total number of photons produced during
the non-equilibrium stages of a ``quenched'' phase transition (which lasts {\em
only for a few} fms./c until the order parameter reaches its equilibrium
value), can become comparable to the number of photons 
released via (perturbative) decay of $all$ the neutral pions produced in the
heavy-ion collision (time scale for decay $\sim 10^7$ fm/c).

In this article we focus our attention on the couplings of the {\em neutral
pion only}, to electromagnetism, through the axial anomaly. We work within the
framework of the linear sigma model and introduce the anomalous interaction of
the $\pi^0$ by means of the effective Wess-Zumino-Witten vertex. As described
above, a DCC corresponds to a rotation of the chiral order parameter in a
coherent domain. Since electromagnetism remains as an unbroken symmetry, such a
rotation must be confined to the $\sigma-\pi^0$ plane. A rotation that takes
the order parameter out of the neutral meson plane homogeneously, will result
in a global breaking of electromagnetism, which is not acceptable. Localized
regions 
in which the charged pion fields obtain (locally) an expectation value (charged
droplets) that average to zero on large volumes cannot be ruled out. However,
the study of their dynamics involves a non-perturbative treatment of large
amplitude inhomogeneous configurations, a very difficult task and beyond the
scope of this article. An initial step towards understanding the
non-equilibrium evolution of {\em small amplitude} inhomogeneous configurations
was presented in\cite{atta}.
 
Minakata and Muller suggested in a recent work \cite{muller} that  a
rotation in the neutral meson plane could indeed be induced in heavy-ion
collisions. These authors argue that the presence of strong electromagnetic
fields in these collisions will give a ``kick'' to the chiral order parameter
in the $\pi^0$ direction, and that this effect is enhanced by the chiral U(1)
anomaly. In this scenario, during the initial stages of the collision the order
parameter acquires a component in the direction of the neutral pion. This is
consistent with the formation of a ``chirally disoriented'' domain. During the
expansion stage this disoriented region cools down and the order parameter
relaxes to the true vacuum configuration, with the field oriented in the
``$\sigma$'' direction.

The goal of this article is to show that if one begins with the above-mentioned
``disoriented'' initial conditions with a large amplitude component in the
$\pi^0$ direction, the DCC relaxation will be accompanied by a coherent burst
of photons which is released via a {\em novel non-perturbative mechanism} of
parametric amplification, along with with the enhanced pion emission resulting
from the non-equilibrium spinodal instabilities.

The essential idea behind this novel mechanism is that the component of the
homogeneous order parameter along the $\pi^0$ direction (the Fourier mode of
the $\pi^0$ field with zero spatial momentum) couples to the electromagnetic
field through the chiral anomaly, and effectively acts much as a time dependent
``mass'' term or ``squeezing'' term for the photon field that leads to the
resonances in the equations of motion.

This non-equilibrium mechanism exhibits the features of unstable bands, and
exponential growth of the photon mode functions, that are characteristic of
parametric resonance. Furthermore, since the amplification and consequently the
enhanced photoproduction will be confined only to certain frequency bands, this
phenomenon translates into very peculiar distributions for the produced
photons. Furthermore, because of the pseudoscalar nature of the interaction,
there is a polarization asymmetry in the produced photons, i.e. photons with
different polarizations are produced in different amounts.  Both features of
this non-perturbative mechanism might be potential signatures of DCC's or
similar non-equilibrium relaxation phenomena during the chiral phase
transition. Parametric amplification of {\em pions} has been conjectured to
play a role in the hadronization stages of the QGP\cite{mrow}.  

In this article we do not incorporate hydrodynamic expansion and treat the
dynamics via the ``quench'' approximation\cite{wilraj,boyan} from an initial state in local thermodynamic equilibrium above $T_c$.  We also study
the evolution of the system from a non-equilibrium initial state at
$T_i=0$ to explicitly show that the qualitative features are robust and not
tied to the ``quench'' scenario. Numerical results of the semiclassical
evolution\cite{randrup} and 
quantum evolution\cite{cooper} with expansion reveal that the qualitative
features, such as pionic amplification and the estimate of the dynamical time
scales are well described by the ``quench'' scenario.

Our aim in this article is to understand the qualitative aspects of this new
mechanism of photoproduction in the simplest of settings, in light of the
possibility that it could be a signature of DCC's in heavy ion collisions. A deeper study of these and other photoproduction mechanisms including boost invariant
expansion will be presented elsewhere \cite{prem}.

This article is organized as follows: in Section II, we introduce the model and
review the approximations used and their validity. Section III is devoted to a
simple derivation of the equations obeyed by the photon mode functions which in
turn indicate the nature of the non-perturbative processes of anomalous
photoproduction. In Section IV we review the numerical study of the equations
of motion, and obtain the photon spectrum for a number of possible initial
conditions, and in Section V we discuss some of the experimental implications
of these results and potential difficulties for its detection. In Section VI we
present our conclusions and future avenues of study. An appendix is devoted to
a detailed derivation of the non-equilibrium Green's functions which are used
for computing the expectation values of the relevant observables.

\section{The model}
In our subsequent investigations in this article we will work with the linear
$\sigma$ model\cite{koch}, and furthermore, will take into consideration the
electromagnetic interactions of the neutral pion only. The interactions of the
charged pions give rise to out of equilibrium processes that produce a single
(direct) photon at lowest order in $\alpha$ through spinodal growth of charged
pion fluctuations; these processes have recently been 
studied in\cite{photop}. On the other hand, the anomalous interactions of the
$\pi^0$ will result in the distinctly different process of diphoton production,
with the 2 emitted photons being polarized perpendicular to each other and will
lead to a polarization asymmetry due to the pseudoscalar nature of the
interaction. The Lagrangian density of the model under consideration is given
by
\begin{eqnarray}
{\cal{L}} &=&\frac{1}{2}\partial_{\mu}\vec{\Phi}\cdot
\partial^{\mu}\vec{\Phi}-\frac{1}{2}m^2(t)\vec{\Phi}\cdot\vec{\Phi}+\lambda 
(\vec{\Phi}\cdot\vec{\Phi})^2-h\sigma 
\label{lagrangian}
\nonumber \\
&-&\frac{1}{4}F_{\mu\nu}F^{\mu\nu}
+\frac{e^2}{32\pi^2}\pi^0F_{\mu\nu}\tilde{F}^{\mu\nu}
\end{eqnarray}
where $\vec{\Phi}$ is an $O(N+1)$ vector, $\vec{\Phi}=(\sigma,\pi^0,\vec{\pi})$
and $\vec{\pi}$ represents the $N-1$ pions,
$\vec{\pi}=(\pi^1,\pi^2,\pi^3,....\pi^{N-1})$ and $m^2(t)$ introduces quench
situation\cite{boyan} ``by hand''. As emphasized in\cite{boyan} the sudden
quench approximation can be generalized to allow a ``cooling'' relaxation time
(of the order of one fm/c) without major qualitative changes in the result and
consistent with the numerical results including expansion\cite{randrup,cooper}.

We then take

\begin{equation}
m^2(t)= \frac{M^2_{\sigma}}{2}\left[ \frac{T^2_i}{T^2_c}\Theta(-t)-1
\right] \; \; \; T_i > T_c.\label{massoft}
\end{equation}

We will also consider below the evolution of the system from a non-equilibrium
initial condition at $T_i=0$ for comparison.

The last term in eq.(\ref{lagrangian}) results from the axial anomaly.  In a
constituent quark model it is known to arise from the quark triangle diagram,
which for 3 flavors of quarks leads to the amplitude for $\pi^0$ decay into two
photons described by the effective interaction vertex in the above
Lagrangian\cite{holstein}.

At this point we want to remark that while the nature of the effective
Lagrangian for the neutral pion decay into photons is clearly understood {\em
in perturbation theory}, it is by no means clear that this effective
interaction also describes strongly out of equilibrium situations. It has been
recently pointed out by Pisarski, Tytgat and Baier et. al.\cite{pisarski} that
{\em in equilibrium} and above the critical temperature there is a drastic
modification of the anomalous vertices.  It is conceivable that the anomalous
decay vertices are modified in strongly out of equilibrium situations as well.
We are currently studying this possibility in detail within the framework of
non-equilibrium field theory in a constituent quark model, but in this article
we accept the anomalous vertex determined by the perturbative triangle diagram
in equilibrium and study its consequences.

The linear sigma model is at best a phenomenological model, and its parameters
are fixed by the low energy phenomenology of pion physics to be\cite{koch}:

\begin{eqnarray}
	M_{\sigma} \approx 600 \mbox{ MeV}     & & \; \; ; \; \;
		f_{\pi} \approx
93 \mbox{ MeV} \; \; ; \; \; \lambda \approx 4.5 \nonumber \\
h \approx (120 \mbox{ MeV})^3             & & \; \; ; \; \; 
	T_c \approx 200 \mbox{ MeV} .
\label{parameters}
\end{eqnarray}

This value of the critical temperature $T_c$ is a consequence of the parameters of the model
and is somewhat larger than the lattice estimates $T_c \approx 150\
\mbox{MeV}$. This is a cutoff theory with an ultraviolet cutoff of the order of
$\Lambda \approx 1\ \mbox{GeV}$.  Although we will be dealing with the $O(4)$
model with 3 pions, we will use the large $N$ limit to provide a consistent,
non-perturbative framework to study the dynamics. For a clear description of
this approximation and its validity, and its applications to a wide variety of
non-equilibrium situations we refer the reader to
\cite{cooper2,strongfields,boyan2}. Eventually one should set $N=3$ which is
the number of physical pions in the real world and though questions may be
raised as to the validity of the large $N$ expansion in this case, it provides
a systematic approximation method to study the non-perturbative physics of the
sigma model.  More importantly perhaps, it is an approximation scheme that
preserves energy conservation (in the absence of expansion), implements PCAC
and satisfies the Ward identities of chiral symmetry. As a result, the pions in
the asymptotic equilibrium state are {\em exactly} massless in the absence of the
explicit symmetry breaking term.

Non-equilibrium quantum field theory requires a path integral representation
along a complex countour in time\cite{schwinger,chou}, with the Lagrangian
density along this contour given by
\begin{equation}
{\cal{L}}_{neq} = {\cal{L}}[\Phi^+]-{\cal{L}}[\Phi^-]
\end{equation}
with the fields $\Phi^{\pm}$ defined along the forward ($+$) and
backward ($-$) time branches. For further details see 
references \cite{cooper2,boyan2}.
The non-equilibrium equations of motion are obtained via the tadpole
method\cite{boyan2}. As mentioned earlier, in the situations of interest to us
in this article, both the $\sigma$  and $\pi^0$ fields may acquire expectation
values. Accordingly, we first shift $\sigma$ and $\pi^0$ by their expectation
values 
in the
non-equilibrium state 
\begin{eqnarray}
\sigma(\vec{x},t)=\phi(t)+\chi(\vec{x},t) \; \; ; \;  \phi(t)=
\langle \sigma(\vec x,t) \rangle \label{split}
\\\nonumber
\pi^0(\vec{x},t)=\zeta(t)+\psi(\vec{x},t) \; \; ; \;  \zeta(t)=
\langle \pi^0(\vec x,t) \rangle 
\end{eqnarray}
with the tadpole conditions
\begin{eqnarray}
\langle\chi(\vec{x},t)\rangle=0 \; \; ; \;\langle\psi(\vec{x},t)\rangle=0  
\; \; ; \;\langle\vec{\pi}(\vec{x},t)\rangle=0 
\label{tadpole}
\end{eqnarray}
being imposed to all orders in the corresponding expansion. To leading order in
$1/N$, the large-$N$ approximation is equivalent to the following Hartree
factorization of the quantum fields:

\begin{eqnarray}
&&\chi^4,\; (\psi^4) \rightarrow 6\langle\chi^2\rangle\chi^2+\text{constant}
,\; (6\langle\psi^2\rangle\psi^2+\text{constant})
\nonumber \\
&&\chi^3, \; (\psi^3) \rightarrow 3\langle\chi^2\rangle\chi,\; 
(3\langle\psi^2\rangle\psi) ,\label{factor}
\\\nonumber
&&(\vec{\pi}\cdot\vec{\pi})^2 \rightarrow (2+\frac{4}{N}) 
\langle{\vec{\pi}}^2\rangle{\vec{\pi}}^2+\text{constant}
\\\nonumber
&&{\vec{\pi}}^2\chi^2,\; ({\vec{\pi}}^2\psi^2) 
\rightarrow {\vec{\pi}}^2\langle\chi^2\rangle+ 
\langle{\vec{\pi}}^2\rangle\chi^2,\; ({\vec{\pi}}^2\langle\psi^2\rangle+ 
\langle{\vec{\pi}}^2\rangle\psi^2) \nonumber \\
&& {\vec{\pi}}^2\chi,\; ({\vec{\pi}}^2\psi) \rightarrow
\langle{\vec{\pi}}^2\rangle\chi, \; (\langle{\vec{\pi}}^2\rangle\psi).  
\end{eqnarray}
In addition to the above factorizations, we will implement a further Hartree
factorization of the Wess-Zumino term,
\begin{equation}
\psi F\tilde{F}\rightarrow \psi\langle F\tilde{F}\rangle.\label{wzwhartree}
\end{equation}
There is no {\em a priori} justification for the factorization of the anomalous
term, but this allows us to treat the full back-reaction problem
self-consistently maintaining energy conservation (in the absence of
expansion).

To leading order in $1/N$, the effective non-equilibrium Lagrangian density is
then given by

\begin{eqnarray}
&&{\cal{L}}[\phi+\chi^+,\vec{\pi}^+]-{\cal{L}}[\phi+\chi^-,
 \vec{\pi}^-]= \left\{\frac{1}{2}(\partial\chi^+)^2+
\frac{1}{2}(\partial\psi^+)^2
+\frac{1}{2}(\partial\vec{\pi}^+)^2 \right. \nonumber \\
&&\left. -{\cal{V}}_1(t)\chi^+-{\cal{V}}_2(t)\psi^+
+8\zeta(t)\phi(t)\chi\psi 
-\frac{1}{2}{\cal{M}_{\chi}}^{+2}(t){\chi}^{+2}
-\frac{1}{2}{\cal{M}_{\psi}}^{+2}(t){\psi}^{+2}
-\frac{1}{2}{\cal{M}_{\vec{\pi}}}^{+2}(t){\vec{\pi}}^{+2}\right\}
\nonumber\\
&&-\frac{1}{4}F^+_{\mu\nu}F^{+\mu\nu}
+\frac{e^2}{32\pi^2}\zeta(t) F^+_{\mu\nu}\tilde{F}^{+\mu\nu}
+\frac{e^2}{32\pi^2}\psi^+\langle F^+_{\mu\nu}\tilde{F}^{+\mu\nu}\rangle
-\left\{+\longrightarrow-\right\}.\label{leff}
\end{eqnarray}
where

\begin{equation}
{\cal{V}}_1 (t)=
\ddot{\phi}(t)+\phi(t)[m^2(t)+4\lambda\phi^2 (t) + 
4\lambda\zeta^2 (t)
+\Sigma(t)]-h \label{vprime}
\end{equation}

\begin{equation}
{\cal{V}}_2 (t)=
\ddot{\zeta}(t)+\zeta(t)[m^2(t)+4\lambda\phi^2 (t) + 
4\lambda\zeta^2 (t)
+\Sigma(t)]\label{vprime2}
\end{equation}

\begin{equation}
{\cal{M}}_{\vec{\pi}}^2(t) = m^2(t)+4\lambda\phi^2 (t) 
+4\lambda\zeta^2 (t)
+ \Sigma(t)\label{pionmass}
\end{equation}
\begin{equation}
{\cal{M}}_{\chi}^2(t) = m^2(t)+12\lambda\phi^2 (t) 
+4\lambda\zeta^2 (t)
+ \Sigma(t).\label{chimass}
\end{equation}

\begin{equation}
{\cal{M}}_{\psi}^2(t) = m^2(t)+12\lambda\zeta^2 (t) 
+4\lambda\phi^2 (t)
+ \Sigma(t) .\label{psimass}
\end{equation}

\begin{equation}
\Sigma(t) = 4\lambda \left [\langle{\vec{\pi}}^2 \rangle(t)-
\langle{\vec{\pi}}^2 \rangle(0)\right] \label{sigma}
\end{equation}
The subtraction of $\langle{\vec{\pi}}^2 \rangle(t)$ at $t=0$ leads to a
renormalization of the mass term given in 
(\ref{massoft})\cite{boyan}.  The presence of the $\chi\psi$ mixing term does
not contribute to the dynamics of the theory in the large $N$ limit because
such a term is formally of ${\cal{O}}(1/N)$ and is therefore subleading. In the
large $N$ limit, the Lagrangian becomes quadratic at the expense of a
self-consistency condition for $\Sigma(t)$.

The pion fields obey a linear Heisenberg evolution equation (in terms of the
self-consistent field), and the Heisenberg field operators can be expanded as

\begin{equation}
\vec{\pi}(\vec{x},t)=\frac{1}{\sqrt{\Omega}}\sum_k\frac{1}{\sqrt{2W_{k,i}}}
[\vec{a}_kU_k(t)e^{i\vec{k}\cdot\vec{x}}+
\vec{a}^{\dagger}_kU_k^*(t)e^{-i\vec{k}\cdot \vec{x}}]
\end{equation}
where $a_k$, $a_k^\dagger$ are the destruction and creation operators of Fock
states associated with the pion field and $\Omega$ is the quantization volume.
The frequencies $W_{k,i}$ determine the initial state and will be specified
below.
 
It is found\cite{cooper,boyan} that the pion mode functions $U_k(t)$ and the
order parameter(s) $\phi(t)$, $\zeta(t)$ obey the following equations of motion
\begin{eqnarray}
&&\ddot{\phi}(t)+[m^2(t)+4\lambda\phi^2(t)+
4\lambda\zeta^2(t)
+\Sigma(t)] \phi(t)-h=0 \label{zeromodeeqn}
\end{eqnarray}

\begin{eqnarray}
&&\ddot{\zeta}(t)+[m^2(t)+4\lambda\phi^2(t)+
4\lambda\zeta^2(t)
+\Sigma(t)] \zeta(t)-\frac{e^2}{32\pi^2}\langle F\tilde{F}\rangle=0 
 \label{zeromodeeqn2}
\end{eqnarray}

\begin{equation}
\left[\frac{d^2}{dt^2}
+k^2+m^2(t)+4\lambda\phi^2(t)+4\lambda\zeta^2(t)+\Sigma(t)
\right]U_k(t)=0\label{modeqns}
\end{equation}

At this stage we must specify the initial conditions on the quantum pion
fields. We will study the evolution with different sets of initial conditions
to determine the sensitivity of the qualitative features of the dynamics to
changes in the initial conditions:

\begin{enumerate}
\item{In the first case we will {\em assume} that at the initial time (the time
of the ``quench'') the quantum fluctuations are in local thermodynamic
equilibrium at the initial temperature $T_i > T_c$ and that the mode functions
$U_k(0)$ correspond to the instantaneous positive frequency modes with mass

\begin{equation}
M^2= m^2(t<0)+ 4 \lambda (\phi^2(0)+ \zeta^2(0)).
\label{inimass}
\end{equation}
In addition, the order parameter is assumed to have a non-zero initial
amplitude in the $\pi^0$ direction i.e. $\zeta(0)\neq0$ and $\phi(0)=0$.}

\item{In the second scenario we assume that the order parameter is initially
displaced from the equilibrium position exactly as in (1), but the modes are
chosen to be at zero temperature, $T_i=0$, so that the
$m^2(t<0)=-\frac{M_\sigma^2}{2}$, and the mass for the modes is
\begin{equation}
M^2= -\frac{M_\sigma^2}{2}+ 4 \lambda (\phi^2(0)+ \zeta^2(0)).
\label{inimass2}
\end{equation}}
\end{enumerate}  

However, in both the above cases, the modes must satisfy,

\begin{equation}
\hspace{0.1in}U_k(0)=1;\hspace{0.1in}\dot{U}_k(0)
=-iW_{k,i} \; \; ; \; \; W_{k,i}=\sqrt{k^2+M^2}. \label{inicon}
\end{equation}

These initial conditions are to be treated only as assumptions that allow
concrete calculations and estimates and are by no means a definitive
description of 
the physics. The correct set of initial conditions should in principle be
obtained by evolving parton cascade models well into the regime in which
hadronic physics can be described by low energy effective lagrangians. This is
certainly an ambitious task and clearly beyond our current understanding. Thus
in view of the lack of clear and precise knowledge of the initial conditions,
we are restricted to investigating a physically plausible initialization of the
dynamics. The above choice is the simplest that allows a concrete calculation
and a quantitative description of the dynamics.

With these initial conditions we are led to the self-consistent
equation\cite{cooper,boyan}

\begin{equation}
\Sigma(t) =(N-1)\int^{\Lambda}\frac{d^3k}{(2\pi)^3}
\frac{|U_k(t)|^2-1}{2W_{k,i}}\coth\left
[\frac {W_{k,i}}{2T_i}\right].\label{fluct}
\end{equation} 

For the initial conditions (2), we must set $T_i=0$.  In addition, the initial
conditions on the order parameters and their derivatives, namely $\phi(0)$,
$\dot{\phi}(0)$, $\zeta(0)$ and $\dot{\zeta}(0)$ must also be specified. At the
moment they are completely arbitrary, constrained only by the requirement that
typical energy densities in a DCC region are $\sim$ a few $\text{GeV/fm}^3$
(which is the typical energy deposited in the central rapidity region in a
heavy-ion collision).

Clearly, a self consistent solution to the above equations will only be
possible after understanding the time evolution of the $\langle
F\tilde{F}\rangle$ term in (\ref{zeromodeeqn2}). This in turn, will require an
investigation of the $\zeta F\tilde{F}$ interaction in (\ref{leff}). As shown
below, it will give rise to out of equilibrium photon production via a process
of parametric resonance.

\section{Photon production}
Let us now turn to the electromagnetic sector of the theory. Since the
processes of interest to us do not involve photons in intermediate states, we
can avoid potential gauge ambiguities by working in the Coulomb gauge and
restricting our attention to the transverse, physical degrees of freedom. In
this gauge, the quadratic part of the Lagrangian for the physical photons, is
given by
 
\begin{eqnarray}
{\cal{L}}^q_{em}=\frac{1}{2}[\partial_\mu\vec{A}_T]^2+\frac{e^2}{32\pi^2}
\frac{\zeta(t)}{f_\pi} F\tilde{F}.
\end{eqnarray}
Since 
\begin{equation}
F\tilde{F}=8(\partial_0A_i)(\partial_jA_k)\epsilon^{0ijk}+ \text{total
divergence}
\end{equation}
${\cal{L}}^{q}_{em}$ may be re-written as,

\begin{eqnarray}
{\cal{L}}^q_{em}&&=\frac{1}{2}[\partial_\mu\vec{A}_T]^2+ \frac{e^2}{4\pi^2}
\frac{\zeta(t)}{f_\pi}(\partial_0A_i)(\partial_jA_k)\epsilon^{0ijk} \\\nonumber
&&=-\frac{1}{2}\vec{A}_T\cdot \Box\vec{A}_T-\frac{e^2}{4\pi^2}
\frac{\dot{\zeta}(t)}{2f_\pi}A^i_T\partial_jA^k_T\epsilon^{ijk} +\text{total
divergence}\label{lagelectro}
\end{eqnarray}
leading to the Heisenberg operator-equation of motion:
\begin{equation}
\Box\vec{A}_T+\frac{e^2}{4\pi^2}
\frac{\dot{\zeta}(t)}{f_\pi}\vec{\nabla}\times\vec{A}_T=0.
\label{emopereq}
\end{equation}

Recall that the interaction term $\psi F\tilde{F}$ has been Hartee-factorized
to give $\psi\langle F\tilde{F}\rangle$ and does not contribute to the
operator equation of motion above.

The physical content of the equations of motion becomes manifest once we
expand the transverse photon field in terms of the circularly polarized states.
For a mode with wave vector $\vec{k}$, consider the triad of vectors
$\hat{k}, \; \vec{\epsilon}_1(\vec k),\; \vec{\epsilon}_2(\vec k)$
with the properties
\begin{eqnarray}
&& \vec{\epsilon}_1(-\vec k) = -\vec{\epsilon}_1(\vec k) \; ; \; 
\vec{\epsilon}_2(-\vec k) = \vec{\epsilon}_2(\vec k) \\ \nonumber
&& \hat k \times \vec{\epsilon}_1(\vec k)= \vec{\epsilon}_2(\vec k)
\; ;  \; \hat k \times \vec{\epsilon}_2(\vec k)= -
\vec{\epsilon}_1(\vec k).
\end{eqnarray}
We can now construct the circular polarization vectors
\begin{equation}
\vec{\epsilon}_+(\vec k) = \frac{i \vec{\epsilon}_1(\vec k)+
\vec{\epsilon}_2(\vec k)}{\sqrt{2}} \; ; \; 
\vec{\epsilon}_-(\vec k)= \frac{i\vec{\epsilon}_1(\vec k)-
\vec{\epsilon}_2(\vec k)}{\sqrt{2}}  
\end{equation}

In terms of these circular polarization vectors we can perform the usual mode
expansion for the fields in terms of annihilation and creation operators:
\begin{eqnarray}
\vec{A}_T(\vec{x},t)&&=\frac{1}{\sqrt{\Omega}}\sum_{\vec{k}}
\vec{A}_T(\vec k , t)\label{emmodeexp}\\\nonumber
&&=\frac{1}{\sqrt{\Omega}}\sum_{\vec{k}}
\frac{1}{\sqrt{2 k}}\left\{
\left[b_+(\vec{k})V_{1\vec{k}}(t)\vec{\epsilon}_+(\vec k) 
+b_-(\vec{k})V_{2\vec{k}}(t)\vec{\epsilon}_-(\vec k)\right] e^{i\vec{k}
\cdot{\vec{x}}}+\text{h.c.}\right\} \label{expansion}
\end{eqnarray}
We find that the mode functions obey the following equations of motion
\begin{eqnarray}
&&\frac{d^2V_{1k}}{dt^2}+k^2V_{1k}(t)-k\frac{e^2}{4\pi^2}
\frac{\dot{\zeta}(t)}{f_\pi}V_{1k}=0
\label{mode1}\\
&&\frac{d^2V_{2k}}{dt^2}+k^2V_{2k}(t)+k\frac{e^2}{4\pi^2}
\frac{\dot{\zeta}(t)}{f_\pi}V_{2k}=0.\label{mode2}
\end{eqnarray}

To these equations of motion we must append initial conditions. Now, we want to
describe the process of emission of physical, circularly polarized massless
photons. Therefore the initial conditions for these mode functions must be:

\begin{eqnarray}
V_{1,2k}(0)=1\; \; \; , \;\dot{V}_{1,2k}(0)=-ik. \label{photoncon}
\end{eqnarray} 

Furthermore, we will assume that at the initial time, the photons are in local
thermodynamic equilibrium at the initial temperature. We will also study
the case in which there are no initial photons by taking the initial
temperature of the photon distribution function to zero.

There are several noteworthy features of these equations of motion that deserve
comment:
\begin{enumerate}
\item{The zero mode of the neutral pion field acts as
time dependent mass term (with a momentum dependence), which will result in
a ``squeezing'' of the quantum fields.}
  
\item{The equations decouple when written in terms of the {\em circular
polarization} mode functions.}
\item{The effective ``mass terms'' for the  two polarizations have opposite
signs. These last two features are a direct consequence 
of the pseudoscalar nature of the coupling of the neutral pion to the
vector field.}
\end{enumerate} 

As the $\pi^0$ zero mode executes quasiperiodic oscillations, the solutions to
the mode equations (\ref{mode1},\ref{mode2}) will display the phenomenon of
parametric resonance, for some values of $k$. In other words, for certain
(forbidden) bands of values in $k$-space, the modes will grow exponentially with
Floquet-like solutions.  This growth of modes will be reflected in the
exponential growth of particle number (in this case, the number of asymptotic
photons). Furthermore, since the coupling affects right and left handed
polarizations differently, there will be a polarization asymmetry in the photon
emission.  A quantitative analysis of this phenomenon will require a solution
of the coupled differential equations (\ref{zeromodeeqn}),
(\ref{zeromodeeqn2}), (\ref{modeqns}), (\ref{fluct}) and (\ref{mode1},
\ref{mode2}) which will be done numerically below.

The number operator for asymptotic photons is given by \cite{photop}:
\begin{equation}
N(k,t)=\frac{1}{2k}\left[
\dot{\vec{A}}_T(\vec{k},t)\cdot\dot{\vec{A}}_T(-\vec{k},t) +k^2
\vec{A}_T(\vec{k},t)\cdot\vec{A}_T(-\vec{k},t) \right]-1 \label{photonnumber}
\end{equation}
In order to obtain the expectation value of this Heisenberg operator in the
initial density matrix, one needs the quantum two-point functions for the
theory. These can be obtained from the expansion (\ref{expansion}).
A detailed derivation of these correlators in terms of non-equilibrium
Green's functions has been given in an appendix. From the expressions for the
Wightman functions (\ref{noneqgf1}) and (\ref{noneqgf2}) it is easy to see that
the expectation value of the number operator in terms of the mode functions
(\ref{mode1},\ref{mode2}), has the following form,

\begin{eqnarray}
\langle N_k(t) \rangle = &&
\frac{1}{4k^2}\coth\left[\frac{k}{2T_i}\right]
\left[|\dot{V}_{1k}(t)|^2+k^2|V_{1k}(t)|^2\right]-\frac{1}{2}+
\nonumber \\
&& \frac{1}{4k^2}\coth\left[\frac{k}{2T_i}\right]
\left[|\dot{V}_{2k}(t)|^2
+k^2|V_{2k}(t)|^2\right]-\frac{1}{2}
\\\nonumber
&&\equiv N_+(k,t)+N_-(k,t)
\end{eqnarray}
Only the symmetric part of the Green's functions contribute to the expectation
value. From this expression for the number of asymptotic, transverse,
massless photons produced, and the evolution equations
(\ref{mode1},\ref{mode2}) it is clear that the number of left and right
circularly polarized photons produced by the evolution of the zero mode of the
neutral pion field will in general be different, since the evolution of the
modes are different.  The polarization asymmetry in the photons produced may be
an experimentally relevant signal. We define this asymmetry as

\begin{equation}
\Xi(k,t)=\frac{(N_+(k,t)-N_-(k,t))}{(N_+(k,t)+N_-(k,t))}.\label{polasymm}
\end{equation}

The dynamics of the process of photon production via the non-equilibrium
evolution of the expectation value of the neutral pion field is then obtained
by solving the self-consistent set of equations of motion
(\ref{zeromodeeqn},\ref{zeromodeeqn2},\ref{modeqns},\ref{fluct},
\ref{mode1},\ref{mode2}) with the initial conditions (\ref{inicon},
\ref{photoncon}).

\section{numerical analysis}
It is convenient to work with dimensionless variables for purposes of numerical
analysis. The relevant dimensionful scale in this problem is $\Lambda_{QCD}
\sim 200\ \text{MeV}=M_F$, and we use it to scale out the dimensions of all the
relevant variables in the equations of motion:
\begin{eqnarray}
&&\xi(t)
=\frac{\zeta(t)}{M_F},\; \; \; \; \tau=M_Ft,\; \; \; \; 
\eta=\frac{\phi}{M_F},\; \; \; \; q=\frac{k}{M_F},
\; \; \; \;\omega_{q}=\sqrt{q^2+\frac{M^2}{M_F^2}} 
\\\nonumber\\
&&{\cal{U}}_q(t)=\frac{U_k(t)}{\sqrt{\omega_q}},\; \; \; \;
{\cal{U}}_q(0)=\frac{1}{\sqrt{\omega_q}},\; \; \; \;
\frac{d{\cal{U}}_q}{d\tau}(0)=-i\sqrt{\omega_q}
\\\nonumber\\
&&\chi_{1,2q}(t)=\frac{V_{1,2k}(t)}{\sqrt{q}},\; \; \; \;
\chi_{1,2q}(0)=\frac{1}{\sqrt{q}},\; \; \; \;
\frac{d{\chi}_{1,2q}}{d\tau}(0)=-i\sqrt{q}
\\\nonumber\\
&&H=\frac{h}{M_F^3}\approx 0.22, \; \; \; \; \; \frac{\Lambda}{M_F}=5,
\; \; \; \; r=\frac{T_i}{M_F}.
\end{eqnarray}
Thus the Wronskian condition for the mode functions defined in terms of
dimensionless variables is
\begin{eqnarray}
\frac{d{\chi}_{1,2q}}{d\tau}\chi_{1,2q}^*-
\frac{{d\chi}_{1,2q}^*}{d\tau}\chi_{1,2q}=-2i.\label{wronskian}
\end{eqnarray}
The expectation value of the anomaly term which appears in the equations of
motion can now be obtained as follows
\begin{eqnarray}
\langle F\tilde{F}\rangle= &&8\langle\partial_0A_T^i\partial_jA_T^k\rangle
\epsilon^{ijk} \\\nonumber =&&8\frac{d}{dt}\int\frac{d^3 k}{(2\pi)^3}
\int\frac{d^3k^\prime} {(2\pi)^3}e^{-i(\vec{k}+\vec{k^\prime})\cdot\vec{x}}
\langle A_T^i(\vec{k},t) A_T^l(\vec{k^\prime},t^\prime)\rangle
(-ik_j^\prime)\epsilon^{ijl}|_{t=t^\prime} .\label{chernsim1}
\end{eqnarray}
The two point correlator in the above expression can be obtained from the
coincidence limit of the Green's functions of the full quantum theory, given in
eqs. (\ref{noneqgf1}) and (\ref{noneqgf2}). In this case, only the
antisymmetric part of the Green's function contributes to the expectation
value.  The expectation of $F\tilde{F}$ in the initial density matrix, is then
found to be:
\begin{equation}
\langle F\tilde{F}\rangle= \frac{M_F^4}{\pi^2}\int
q^3dq\coth\left[\frac{q}{2r}\right]\frac{d}{d\tau}\left(|\chi_{2q}|^2-
|\chi_{1q}|^2\right) \label{chernsim}
\end{equation}
where $q$ is the dimensionless momentum and use was made of the Wronskian
condition (\ref{wronskian}).

The full dynamics of the system can now be followed by numerically solving the
following set of equations:
\begin{eqnarray}
&&\frac{d^2{\eta}(\tau)}{d\tau^2}+\frac{m^2(\tau)}{M_F^2}+4\lambda
\eta(\tau) [\eta^2(t)+ \xi^2(\tau)+\Sigma(\tau)]
-H=0,
\\\nonumber
\\\nonumber
&&\frac{d^2\xi(\tau)}{d\tau^2}+\frac{m^2(t)}{M_F^2}\xi(\tau)
+4\lambda\xi(\tau)\left[\eta^2(\tau)+\xi^2(\tau)+\Sigma(\tau)\right]
-\Gamma(\tau)=0
\\\nonumber
\\\nonumber
&&\frac{d^2{\cal{U}}_q}{d\tau^2}
+\left(q^2+\frac{m^2(\tau)}{M_F^2}+4\lambda\eta^2(t)+4\lambda\xi^2(t)
+4\lambda\Sigma(\tau)\right){\cal{U}}_q(t)=0
\\\nonumber
\\\nonumber
&&\frac{d^2\chi_{1q}}{d\tau^2}+\left(q^2-\frac{\alpha M_F}{\pi f_\pi} 
q\frac{d\xi}{d\tau}\right)\chi_{1q}=0
\\\nonumber
\\\nonumber
&&\frac{d^2\chi_{2q}}{d\tau^2}+\left(q^2+\frac{\alpha M_F}{\pi f_\pi} 
q\frac{d\xi}{d\tau}\right)\chi_{2q}=0
\\\nonumber
\\\nonumber
\end{eqnarray}

with the initial conditions

\begin{eqnarray}
&&\hspace{0.1in}{\cal{U}}_q(0)=\frac{1}{\sqrt{\omega_q}}, \hspace{0.1in}
\dot{\cal{U}}_q(0)
=-i\sqrt{\omega_q}
\\\nonumber
\\\nonumber
&&\chi_{1,2q}(0)=\frac{1}{\sqrt{q}}\; \; \; ,  \; \dot{\chi}_{1,2q}(0)=-i\sqrt{q}
\label{fininicond}
\end{eqnarray}

and the self-consistent fluctuations given by

\begin{eqnarray}
&&\Sigma(\tau)=\frac{2}{4\pi^2}\int_0^{5} q^2dq
(|{\cal{U}}_q(\tau)|^2-|{\cal{U}}_q(0)|^2)\coth
\left[\frac{\omega_q}{2r}\right]
\\\nonumber
\\\nonumber
&&\Gamma(\tau)=\frac{\alpha M_F}{8\pi^3f_\pi}\int^{5}_0
q^3dq\coth\left[\frac{q}{2r}\right]\frac{d}{d\tau}\left(|\chi_{2q}|^2-
|\chi_{1q}|^2\right) \label{selfconsfluc}
\end{eqnarray}
The number of photons per unit volume per unit phase space is then
\begin{eqnarray}
&&(2\pi)^3\ \frac{d(\langle N_q(\tau)\rangle -\langle
N_q(0)\rangle)/\Omega}{d^3q}
\equiv \frac{n(q,\tau)}{|q|}
\label{numnumb}\\\nonumber
&&=\frac{1}{4q}\coth\left[\frac{q}{2r}\right]\left[
|\frac{d\chi_{1q}}{d\tau}|^2+|\frac{d\chi_{2q}}{d\tau}|^2
+q^2\left(|\chi_{1q}|^2+|\chi_{2q}|^2\right)-4q\right]
\end{eqnarray}
Here, we have taken the continuum limit, and subtracted out the number of
photons present at $t=0$ which forms a thermal background at the initial
temperature $T_i$. We have also introduced the quantity $n(q,\tau)=
(2\pi)^3\ |q|\frac{d(\langle N_q(\tau)\rangle -\langle
N_q(0)\rangle)/\Omega}{d^3q}$. Our numerical results for the photon spectrum
and the invariant rate will be expressed
in terms  of this quantity. The invariant rate for photon production can be
obtained by 
looking at the time derivative of (\ref{numnumb}) and using the mode equations,
it is found to be:
\begin{equation}
(2\pi^3)\frac{d(\langle\dot{N}(\tau)\rangle/\Omega)}{d^3q}=
\frac{\alpha M_F}{4\pi f_{\pi}}
\dot{\xi}\coth[\frac{q}{2r}]
\frac{d}{d\tau}\left[|\chi_{1q}|^2-|\chi_{2q}|^2
\right].
\end{equation}
A comparison of this expression for the rate and that for the expectation value
of the anomaly term in eq.(\ref{chernsim}) clearly indicates the the important
role played by the latter quantity in ensuring energy conservation. The
photons 
produced lead to a non-zero expectation value for $F\tilde{F}$ which then feeds
back into the equation of motion for the zero mode.

The above expression for the rate will also turn out to be useful for obtaining
 quantitative estimates of the location of the unstable bands.  In
equilibrium, $|\chi_{1q}|^2=|\chi_{2q}|^2=1/q$, and thus the rate
vanishes. However, out of equilibrium and particularly for the processes that
we have considered here, both $|\chi_{1q}|$ and $|\chi_{2q}|$ will
acquire small time dependent modulations around a mean value, with a frequency
that is roughly $\sim q$, since that is the relevant scale in the problem.
Thus the oscillation frequency of
$\frac{d}{d\tau}(|\chi_{1q}|^2-|\chi_{2q}|^2)$ is $\sim 2q$. Now the maximum
buildup of particles will occur when this oscillation is in phase with the
derivative of the zero mode $\dot{\xi}(\tau)$, resulting in a positive rate for
all times, causing a resonant production of photons. Therefore the unstable
bands will be determined by the condition
\begin{equation}
q_{res}M_F=k_{res}\approx
\frac{\omega_{\pi^0}}{2}.\label{resoncond}
\end{equation}
with $\omega_{\pi^0}$ being the oscillation frequency of the expectation value
of the $\pi^0$ component of the order parameter.  For small oscillations in the
$\pi^0$ direction, $\omega_{\pi^0}=M_\pi=140$ MeV and so the resonant band
should be at $k=70$ MeV, which is exactly the energy of two back to back
photons produced by a $\pi^0$ at rest. The validity of these qualitative
arguments and particularly that of eq. (\ref{resoncond}) will be borne out by
the results of the numerical analysis which we present below.

For all the cases that we considered, we chose to represent the ``quench'' from
an initial temperature $T_i=1.1T_c=220$ MeV to zero temperature, and we tracked
the evolution of the system up to a time of about $10$ fm/c which is the typical
time scale after which the effects of hydrodynamical expansion become
important.  Since we are interested in studying the optimal situation for large
anomalous photon production from the non-equilibrium evolution of the neutral
pion, we have performed our numerical analysis in all cases with the initial
value of the $\sigma$ component as well as its initial time derivative being
zero. An initial condition with both the neutral pion and $\sigma$ components
being different from zero will result in a smaller photon yield. This is
because, for a given initial energy density, if the $\sigma$ component is
large, the $\pi^0$ component must be smaller and it is the $\pi^0$ zero mode
that drives the parametric amplification.  Furthermore since the sigma model is
an effective theory below 1 GeV we look only at photons with momenta less than
this value. Photons with energy larger than a GeV will be indinstinguishable
from the hard photon background and much less significant as a signature of low
$p_T$ physics associated with DCC's or the chiral phase transition.

{\bf{Results:}}

The results of the numerical calculations are shown in figures (1) through (6).
We first look at the result obtained for small oscillations in the $\pi^0$
direction with initial conditions $\zeta(0)=\xi(0)M_F=0.1M_F$,
$\dot{\xi}(0)=0$, $\phi(0)=\dot{\phi}(0)=0$. This is the perturbative
regime, in which $\xi(\tau)$ undergoes small amplitude (almost undamped)
oscillations, with emission of back to back photons at 70 MeV.  Fig.(1) shows
the spectrum of photons emitted per unit volume,
$n(q,\tau)=(2\pi)^3|q|\frac{d(N(\tau)-N(0))/\Omega}{d^3q}$, (in units of
$\text{fm}^{-1}$) at time $t=10$ fm/c, for the small amplitude initial
conditions described above. Here, we have subtracted out the thermal
distribution of photons that are present initially at $t=0$ when the system is
in equilibrium at a temperaure, $T_i=220$ MeV.  Clearly, there is a resonant
peak which upon closer inspection is found to be exactly centered around $70$
MeV. This corresponds to the usual perturbative decay of zero momentum pions
and as expected, the peak value of the photon number, $2.75\times10^{-5}$, is
${\cal{O}}(\alpha^2)$. In this perturbative (small amplitude) regime the
location of the unstable band for small oscillations can be easily obtained by
solving the mode equations (\ref{mode1},\ref{mode2}) assuming a sinusoidal time
dependence for the zero mode, which leads to a Mathieu equation for the
non-zero modes for which the position of the unstable bands can be found in
textbooks\cite{whitaker}.

Figs.(2.a) and (2.b) show $\dot{\xi}(\tau)$ and
$n(q,\tau)=(2\pi)^3|q|\frac{d(N(\tau)-N(0))/\Omega}{d^3q}$, (in units of
$\text{fm}^{-1}$) at time $t=10$ fm/c respectively, for the initial conditions
$\zeta(0)=\xi(0)M_F=1.0 M_F$, $\dot{\xi}(0)=0$,
$\phi(0)=\dot{\phi}(0)=0$. This corresponds to preparing a configuration
with an energy density $\lambda(\zeta(0)^2-f_\pi^2)^2\approx0.5$
$\text{GeV/fm}^3$ above the thermal energy density.  $\dot{\xi}$ now evolves
with a larger amplitude and also a higher frequency as shown in Figure
(2.a). The small damping has its origins in mode mixing and excitations of the
quantum fluctuations (pions and ``sigmas'') via spinodal and parametric
amplification, and also to a smaller extent, the contribution from photon
production. The spectrum of emitted photons is shown in Fig.(2.b). It features
a prominent peak centered around $306$ MeV with a much larger value, 0.005
$\text{fm}^{-1}$ (vs. 0.0000275 $\text{fm}^{-1}$ in Fig.(1)) which is of
${\cal{O}}(\alpha)$ (rather than ${\cal{O}}(\alpha^2)$ as in Fig.(1)) showing
explicitly the non-perturbative nature of the amplification phenomenon. The
band is also quite broad with a width $\approx 120$ MeV.  This is clearly a
non-perturbative effect caused by parametric resonance. In fact our qualitative
argument (\ref{resoncond}) for obtaining the resonant frequency works rather
well here. $\omega_{\pi^0}$ can be read off from the oscillations of the zero
mode in Figure (2.a) and is $\approx\frac{2\pi}{2}\times 200$ MeV so that
$k_{res}\approx\omega_{\pi^0}/2\approx 310$ MeV. To illustrate the point of
parametric amplification, Fig.(2.c) shows the invariant rate
$\dot{n}(q,\tau)=(2\pi)^3|q|\frac{d\dot{N}(\tau)/ \Omega}{d^3q}$, (in units of
$\text{fm}^{-2}$) for $q$ at the center of the resonant band $k=k_{res}$ as a
function of time (in units of fm/c).  It is important to remark that unlike the
small amplitude regime, in this large amplitude case, the rate is
${\cal{O}}(\alpha)$ rather than the perturbative result ${\cal{O}}(\alpha^2)$.
This figure clearly displays a {\em positive} quasiperiodic solution with a
growing envelope (the ``true'' invariant rate should be obtained by dividing by
the total number of photons as a function of time, as the above rate has been
defined as the time derivative of the number of photons, not of its
logarithm). Our numerical studies show that, away from the center of the
resonant band, the rate samples negative values also, thus leading to a lower
photon yield for $q$ values away from the center.

With a larger initial amplitude for the $\pi^0$ zero mode, we see the same
non-perturbative effects as before with a much larger enhancement. Figs.(3.a),
(3.b) and (3.c) show the results for initial conditions
$\zeta(0)=\xi(0)M_F=2.0M_F$, $\dot{\xi}(0)=0$,
$\phi(0)=\dot{\phi}(0)=0$. The energy density corresponding to this initial
state is $\approx 12$ $\text{GeV/fm}^3$. It is not surprising that for such a
large energy density, the peak in the photon spectrum (Fig.(3.b)) and the
invariant rate at the resonant frequency (Fig.(3.c)) are about an order of
magnitude larger than the previous case. Furthermore the position of the peak
is now shifted further to the right to $625$ MeV in qualitative agreement with
the argument (\ref{resoncond}).

As anticipated in the discussions of the previous section, these
non-perturbative effects and the pseudoscalar nature of the interaction, will
give rise to a polarization asymmetry in the photons produced, as well as a
non-equilibrium expectation value for $F\tilde{F}$.
 
In Figures (4.a) and (4.b) we show the polarization asymmetry at the resonant
value of $q$;
$\Xi(q_{res},\tau)
=(n_+(q_{res},\tau)-n_-(q_{res},\tau))/(n_+(q_{res},\tau)+n_-(q_{res},\tau))$
as a function of time for $k=306$ $\text{MeV}$, $\xi(0)=1.0$ (corresponding to
the initial conditions of Figs.(2)) and $k=625\ \text{MeV}$, $\xi(0)=2.0$
(corresponding to Figs.(3)). Although this asymmetry is a (quasi) oscillatory
function of time, the average over a time scale of about 10 fm/c, yields an
estimate of $|\Xi(q_{res},t)|=0.05-0.10$ for the values used in the figures.

The behaviour of $\Gamma(\tau)=\alpha M_F\langle F\tilde{F}\rangle/8\pi f_{\pi}
$ (in units of $M^4_F$)
as a function of time up to $10$ fm/c  is shown in Figures (5.a) and
(5.b). Though the behaviour is 
oscillatory, it is not symmetric about zero and is of order $10^{-4}-10^{-3}$
 $M_F^4$. Thus $\langle F\tilde{F}\rangle$ itself is of order $0.1-1$ $M_F^4$.

We have also studied non-equilibrium initial conditions that do not
correspond to local thermodynamic equilibrium by setting $T_i$ to zero, in the
initial conditions and relevant quantities. 

Fig.(6.a) shows $n(q,\tau)=(2\pi)^3|q|\frac{d (N(\tau)-N(0))/\Omega}{d^3q}$,
(in units of 
$\text{fm}^{-1}$) at time $t=10$ fm/c respectively, for the  
initial conditions $\zeta(0)=\xi(0)M_F=1.0 M_F$, $\dot{\xi}(0)=0$, 
$\phi(0)=\dot{\phi}(0)=0$, $T_i=0$. Comparing this figure with
Fig.(2.b) we see that the main features are qualitatively the same.
The amplitude is smaller in Fig.(6.a), reflecting the Bose-enhancement for
$T_i\neq 0$ 
in Fig.(2.b). There is also a slight shift in the center of the
unstable band. In Fig.(6.a) the peak is at $k=284 \text{MeV}$ rather
than at $k=306 \text{MeV}$ as in Fig.(2.b). Fig.(6.b) shows
the invariant rate $\dot{n}(q,\tau)=(2\pi)^3|q|\frac{d\dot{N}(t)/
\Omega}{d^3q}$, (in units of $\text{fm}^{-2}$) for $q$ at the center of the
resonant band $k=k_{res}$
as a function of time (in units of fm/c) for the same initial
conditions as for Fig.(6.a). Fig.(6.c) shows the polarization asymmetry for
the same initial conditions. 

Clearly the effect is fairly robust in that
it mainly depends on the non-equilibrium aspects of the evolution with
a rather small quantitative change due to Bose-enhancement if the initial
state is in LTE at $T_i\neq0$. A similar conclusion is obtained after
numerical 
analysis with the initial conditions of Figs.(3.a-c) and $T_i=0$,
with the same qualitative features modified slightly by Bose-enhancement (or
the lack thereof).

\section{Possible Experimental Signatures:}

In a heavy-ion collision one of the most important contributions to the  
production  of low energy photons is from the anomalous single particle decay
of the neutral pion. The neutral pions formed during the hadronization and
chiral symmetry breaking stages will mainly decay at freeze out, since their
lifetimes (at rest) are  $\approx 10^7$ fm/c. 

The mechanisms studied above are effective {\em only} during the typical time
scales for the non-equilibrium effects and for large initial
amplitudes of 
the neutral pion component of the order parameter. 
 Numerical estimates of the 
expansion time scales in the semiclassical\cite{randrup} and 
quantum large $N$ approximations within the sigma model\cite{cooper} reveal
a longitudinal expansion time scale between 5 to 10 fm/c (spherical
expansion scales are somewhat shorter). Although we have not incorporated
expansion in our studies, it is a plausible assumption that even if the initial
amplitude for the neutral pion component is large, giving rise to the
non-perturbative parametric amplification phenomena described above, this
amplitude will have diminished substantially by the end of the expansion
stage. Thus 
most of the non-equilibrium, non-perturbative phenomena will take place
{\em during} the initial stages for a time-scale of about 10 fm/c. 
Only the photons produced during this non-equilibrium stage will feature
the abnormal distributions studied above, whereas the photons
 emitted via the perturbative (small amplitude) decay of the neutral pion after
freeze out will have the typical distribution with a peak at 70 MeV for
decay at rest. 

Therefore an experimentally  meaningful quantity  is the ratio of
the total number of photons produced during the non-equilibrium stages with
abnormal (parametrically amplified) distributions 
 to half the initial number of $\pi^0$'s. Notice that this {\em assumes} that
{\em all}
the $\pi^0$'s produced will decay perturbatively to 2 photons. Thus this ratio
gives a conservative estimate of the experimental relevance of the 
abnormal electromagnetic signal
and provides an estimate for the experimental signature of the
non-equilibrium aspects of anomalous photon production during the chiral
phase transition.

 We found numerically that the pion number density produced via parametric
amplification and spinodal instabilities during the non-equilibrium
stages, including the pions present in the thermal bath is about
$0.6/\text{fm}^3$ for $\zeta(0)=\xi(0)M_F=1.0M_F$; and  
$\approx 5/\text{fm}^3$ for $\zeta(0)=\xi(0)M_F=2.0M_F$, where $M_F=200$ MeV. 
The corresponding numbers for the photons produced by parametric amplification
out of equilibrium are 
$0.0015/\text{fm}^3$ and $0.005/\text{fm}^3$ respectively.

 Thus the ratio of the number of
amplified photons to the number of decay photons is $\approx 0.1 -0.2 \%$, 
 a rather small number. 
If the detector can measure final state polarization of photons, this
signal must be correlated with the total (integrated over the lifetime of the
``fireball'') polarization asymmetry  
$\Xi_{tot} = |n_+-n_-|/(n_+ + n_-) \approx 0.05-0.1$ where the sum
in the denominator refers to the parametrically amplified photons.

Therefore the experimental signature of these non-equilibrium effects resulting
in an enhancement of photons produced through the non-equilibrium decay of the
neutral pion, will be a peak at momenta $k \geq 300 $ MeV with an amplitude
less than a percent of the amplitude of the usual $\pi^0$ peak and with a
polarization asymmetry between $5-10 \%$. (Recall, however that we have assumed
that all the neutral pions produced will contribute to the $70$ MeV peak by
decaying at rest, to two photons.)  The photon energy corresponding to the
center of the peak increases with the initial energy density at the onset of
the non-equilibrium stage.

Clearly these are small effects leading to less than 1 percent signal over
background relative to the usual $\pi^0$ peak. 
However, our estimate is only a lower bound and the effect could still be
within the limits of resolution of the detectors and could potentially
provide  one of the electromagnetic signatures of DCC's.  

In the energy region $k \geq 280\ \text{MeV}$ (corresponding to an initial
value of $\langle \pi^0 \rangle \approx 200\ \text{MeV}$) the most problematic
interference for detection is the decay of the $\eta$ pseudoscalar meson into
two photons, with a branching ratio of about $40 \%$. This neutral mode of
decay for the $\eta$ meson has the same features as that of the neutral pion,
insofar as angular distribution and polarization asymmetry are concerned. We
can provide a 
rough (albeit naive) estimate of the number of diphotons produced by $\eta$
decay by assuming that the $\eta$'s are in equilibrium at the initial
temperature since they would not undergo parametric or spinodal amplification;
this yields an $\eta$ number density of about $0.088/\text{fm}^{3}$. This in
turn leads to about $0.035/\text{fm}^{3}$ di-photons at an energy $\approx 275$
 MeV. Thus if the initial condition corresponds to a disorientation in
the $\pi^0$ direction with an expectation value of the neutral pion field
$\approx 200$ MeV, this would translate to a peak in the photon spectrum in the
vicinity of the peak associated with the $\eta$ meson with about $5 \%$ ratio
of the peak values. Clearly this is {\em not} a generic initial condition; for
larger values of the $\pi^0$ component, the peaks in the distribution will move
to larger values of the momenta and their amplitude will be enhanced. Although
one could argue that for larger values of the momenta the background photon
production from higher mass resonance decays will be important, these will be
suppressed by thermal factors and smaller branching ratios.

\section{Conclusions and further questions}

In this article we have focussed on the study of photon production enhancement
through the non-equilibrium stages of relaxation of a DCC via the $U_A(1)$
anomaly.

The premise of the study is that if during a heavy-ion collision a state of
large energy density is formed in which the chiral order parameter is
``disoriented'' with a large amplitude in the $\pi^0$ direction, such a state
will relax via parametric and spinodal amplifications and will also lead to an
enhanced production of photons through the $U_A(1)$ anomaly.

We have studied the optimal situation in which the order parameter has the
largest amplitude along the neutral pion direction, with typical amplitudes for
the zero mode of the neutral pion $\langle \pi^0 \rangle = 200-400$ MeV
corresponding to energy densities in the initial state between $1-10\
\text{GeV}/\text{fm}^3$.  We have followed the evolution during a ``quenched''
phase transition from an initial state in local thermodynamic equilibrium at a
temperature slightly above the critical temperature, cooled instantaneously to zero temperature, as well as from a non-equilibrium  initial condition
at $T_i=0$.

We found that the oscillations of the large amplitude neutral
pion lead to parametric amplification of circularly polarized
photons resulting in unstable $k$-bands within which the photon mode
functions grow almost exponentially leading to a distinct distribution
of the produced photons. The peak of the distribution is correlated with
the initial amplitude of the neutral pion component of the order 
parameter and  the initial energy density. This is a non-perturbative
phenomenon as clearly seen in the ratio of the amplitudes of the 
photon distribution functions at the peak between the small 
amplitude (perturbative) and the large amplitude (non-perturbative)
regime. 

For initial conditions in which the order parameter has a component only along
the $\pi_0$ direction, we found that the peak in the photon spectrum moves
continuously with the initial amplitude of the $\pi^0$, from $k \approx 300$
MeV for $\langle \pi^0 \rangle = 200$ MeV corresponding to an initial energy
density $\approx 1$ $\text{GeV}/ \text{fm}^3$, to $k \approx 620$ MeV for
$\langle \pi^0 \rangle = 400$ MeV corresponding to an initial energy density
$\approx 12$ $\text{GeV}/ \text{fm}^3$ during a time scale of 10 fm/c. The
pseudoscalar nature of the interaction results in a net polarization asymmetry
of the final state photons which ranges between $0.05-0.1$ for the above
initial conditions. Both the total number of photons created during this 
process and their polarization asymmetry are correlated with a non-equilibrium
expectation value of the anomaly term $F\tilde{F}$.

These novel distributions for photons produced by the ``anomalous''
decay result in peaks in the photon spectrum whose ratio to the normal
peak resulting from the ``perturbative'' neutral pion decay at 70 MeV is
less than the one percent level. Experimentally this is a rather small signal,
but perhaps when correlated with the polarization asymmetry, it may result
in an unambiguous signal for non-equilibrium processes during the
chiral phase transition. 

We believe that these results are fairly encouraging and justify a deeper study
of these ``anomalous non-equilibrium effects''. The next step in the program is
to extend the investigation of reference\cite{pisarski} to understand the
effective anomalous vertex out of equilibrium.  We plan to study this, within
the framework of a constituent quark model or alternatively a NJL model in
which the triangle diagram out of equilibrium should furnish the effective
vertex for neutral pion decay.  After a thorough understanding of the effective
$\pi^0\rightarrow2\gamma$ vertex and its translation to an effective mesonic
theory, we will include hydrodynamic expansion by describing the dynamics in
terms of boost invariant variables as in\cite{cooper}.

\acknowledgements 
D.B. and S.P.K. thank B. Mueller for enlightening
discussions.  D.B. thanks J. D. Bjorken, R. Pisarski, J. Randrup, V. Koch,
X. N. Wang, E. Engels and R. Willey for illuminating conversations and
suggestions and acknowledges support from N.S.F. through Grant: PHY-9302534.
R.H. and S.P.K were supported in part by DOE grant $\#$ DE-FG02-91-ER40682.

\appendix

\section{Derivation of Non-Equilibrium Green's functions:}

In this appendix we present a complete derivation of the non-equilibrium
Green's functions which are needed primarily for the computation of the
expectation values of observables, such as the photon number density, and
$F\tilde{F}$.

The Lagrangian density given by eqn. (\ref{lagelectro})
can be written as
\begin{equation}
-\frac{1}{2}A^m_T\left[\delta^{mn}\Box+\frac{e^2}{4\pi^2}
\frac{\dot{\zeta}(t)}{f_\pi}\partial_j\epsilon^{mjn}\right]A_T^n
+\text{total divergence}.
\end{equation}
We now define the Fourier transform of the vector potential as,

\begin{equation}
A_T^m(\vec{x})=\int \frac{d^3k}{(2\pi)^3}e^{-i\vec{k}\cdot\vec{x}}
A_T^m(\vec{k}).
\end{equation}
Therefore

\begin{equation}
S_{em}=\int dt \frac{d^3k}{(2\pi)^3}\left\{-\frac{1}{2}
A_T^m(\vec{k})\left[\delta^{mn}(\frac{d^2}{dt^2}+k^2)
+i\frac{e^2}{4\pi^2}\frac{\dot{\zeta}(t)}{f_\pi}\epsilon^{mjn}k_j\right]
A_T^{*n}(\vec{k})\right\}.
\end{equation}
It also turns out to be convenient to express all our equations in terms of
the two operators,

\begin{equation}
P^{mn}=(\delta^{mn}-\frac{k^mk^n}{k^2}), \; \; \; \; \; 
f^{mn}=\epsilon^{mjn}\frac{k^j}{k},
\label{operatordef}
\end{equation}
where $k=\sqrt{k^jk^j}$ and $P^{mn}$ is the projector onto transverse
states. It 
can be easily verified that these operators have the following properties:

\begin{equation}
P^{mn}f^{nl}=f^{ml},\; \; \; \; \; \;f^{mn}f^{nl}=-P^{ml}.
\end{equation}
The action for the electromagnetic sector is then,

\begin{eqnarray}
S_{em}&&=\int dt \frac{d^3k}{(2\pi)^3}\left\{-\frac{1}{2}
A^i_T(\vec{k})P^{im}\left[\delta^{mn}(\frac{d^2}{dt^2}+k^2)
+ik\frac{e^2}{4\pi^2}\frac{\dot{\zeta}(t)}{f_\pi}f^{mn}\right]
P^{jn}A^{*j}_T(\vec{k})\right\}\label{emlag}\\\nonumber
&&=\int dt \frac{d^3k}{(2\pi)^3}\left\{-\frac{1}{2}
A^i_T(\vec{k})\left[P^{ij}(\frac{d^2}{dt^2}+k^2)
+ik\frac{e^2}{4\pi^2}\frac{\dot{\zeta}(t)}{f_\pi}f^{ij}\right]
A^{*j}_T(\vec{k})\right\}.\\\nonumber
\end{eqnarray}
This is simply a quadratic action for two coupled complex scalar fields. The
time dependent mixing term, driven by the coherent oscillations of 
the neutral pion zero mode $\zeta(t)$, acts like a squeeze parameter. One can
decouple the fields, either at the level of the action by diagonalizing the
quadratic form, or by rewriting the equations of motion in terms of the
appropriate degrees of freedom. We will adopt the latter method. In any case,
the problem is completely solved once we obtain all the relevant
non-equlibrium two-point functions of the theory. 
The Green's functions ${\cal{G}}_{ij}$ for the operator
appearing 
in the 
quadratic action must satisfy:

\begin{eqnarray}
{\cal{D}}_{ij}{\cal{G}}_{jk}(t,t^\prime)=-P_{ik}\delta(t-t^\prime)
\label{gfunction}
\end{eqnarray}
where

\begin{eqnarray}
{\cal{D}}_{ij}=\left\{P_{ij}(\frac{d^2}{dt^2}+k^2)+ik\frac{e^2}{4\pi^2}f_{ij}
\frac{\dot{\zeta}(t)}{f_\pi}\right\}.
\end{eqnarray}
The Green's function can be expressed in terms of the two linearly independent
operators defined in (\ref{operatordef}) as

\begin{eqnarray}
{\cal{G}}_{jk}(t,t^\prime)=C(t,t^\prime)P_{jk}+D(t,t^\prime)f_{jk}.
\end{eqnarray}
Substituting this into eq. (\ref{gfunction}), and equating the symmetric
and antisymmetric parts of the two sides of the equation, we find:

\begin{eqnarray}
&&\frac{d^2C}{dt^2}+k^2C-ik\frac{e^2}{4\pi^2}\frac{\dot{\zeta}(t)}{f_\pi}D=
-\delta(t-t^\prime)\\\nonumber
&&\frac{d^2D}{dt^2}+k^2D+ik\frac{e^2}{4\pi^2}\frac{\dot{\zeta}(t)}{f_\pi}C=0.
\end{eqnarray}
These two differential equations can be decoupled by defining the functions

\begin{equation}
G_1(t,t^\prime)=C+iD,\; \; \; \; \;G_2(t,t^\prime)=C-iD,
\end{equation}
which then yields

\begin{eqnarray}
&&\frac{d^2G_1}{dt^2}+k^2G_1-k\frac{e^2}{4\pi^2}\frac{\dot{\zeta}(t)}{f_\pi}
G_1=-\delta(t-t^\prime)\label{G1andG2}\\\nonumber
&&\frac{d^2G_2}{dt^2}+k^2G_2+k\frac{e^2}{4\pi^2}\frac{\dot{\zeta}(t)}{f_\pi}
G_2=-\delta(t-t^\prime).
\end{eqnarray}
Also,

\begin{equation}
{\cal{G}}_{jk}(t,t^\prime)=\left[\frac{G_1+G_2}{2}P_{jk}
+\frac{G_1-G_2}{2i}f_{jk}\right].
\end{equation}

{\noindent}$G_1$ and $G_2$ now satisfy the simple Green's function 
equations (\ref{G1andG2})
 and can be readily
expanded in terms of the mode functions (\ref{mode1},\ref{mode2}) as,

\begin{eqnarray}
G_{1,2}(t,t^\prime)=&&\left[A^>_{1,2}V_{1,2}(t)V_{1,2}^*(t^\prime)
+B^>_{1,2}V_{1,2}^*(t)V_{1,2}(t^\prime) \right]\Theta(t-t^\prime)
\\\nonumber
&& +\left[A^<_{1,2}V_{1,2}^*(t)V_{1,2}(t^\prime)
+B^<_{1,2}V_{1,2}(t)V_{1,2}^*(t^\prime) \right]\Theta(t^\prime-t)
\\\nonumber
\end{eqnarray}

where the mode functions must satisfy the homogeneous equations:
\begin{eqnarray}
&&\frac{d^2V_{1k}}{dt^2}+k^2V_{1k}(t)-k\frac{e^2}{4\pi^2}
\frac{\dot{\zeta}(t)}{f_\pi}V_{1k}=0
\\
&&\frac{d^2V_{2k}}{dt^2}+k^2V_{2k}(t)+k\frac{e^2}{4\pi^2}
\frac{\dot{\zeta}(t)}{f_\pi}V_{2k}=0.
\end{eqnarray}

The boundary conditions on the Green's functions, which include
conditions of 
continuity and the KMS periodicity condition at finite initial
temperature, allow all
the constants -- namely, the A's and the B's to be uniquely determined, finally
yielding

\begin{eqnarray}
{\cal{G}}_{jk}(t,t^\prime)={\cal{G}}_{jk}^{>}(t,t^\prime)\Theta(t-t^\prime)
+{\cal{G}}_{jk}^{<}(t,t^\prime)\Theta(t^\prime-t)
\end{eqnarray}

with the Wightman functions

\begin{eqnarray}
{\cal{G}}_{jk}^{>}(t,t^\prime)
&&=i\langle A^{-}_{Tj}(\vec{k},t)A^{+}_{Tk}(-\vec{k},t^\prime)\rangle
\label{noneqgf1}\\\nonumber\\\nonumber
&&=\left[\frac{P_{jk}}{2}\frac{1}{2ik}\left\{
(1+n_k)V_{1k}(t)V^*_{1k}(t^\prime)
+n_kV_{1k}^*(t)V_{1k}(t^\prime)\right.\right.\\\nonumber\\\nonumber
&&\left.\left.
\hspace{.5 in}+(1+n_k)V_{2k}(t)V_{2k}^*
(t^\prime)+n_kV_{2k}^*(t)V_{2k}(t^\prime)
\right\}\right.
\\\nonumber\\\nonumber
&&\left.
-\frac{f_{jk}}{2}\frac{1}{2k}\left\{(1+n_k)V_{1k}(t)V^*_{1k}(t^\prime)
+n_kV_{1k}^*(t)V_{1k}(t^\prime)\right.\right.\\\nonumber\\\nonumber
&&\left.\left.
\hspace{.5 in}-(1+n_k)V_{2k}(t)V_{2k}^*
(t^\prime)-n_kV_{2k}^*(t)V_{2k}(t^\prime)
\right\}\right]
\end{eqnarray}
and

\begin{eqnarray}
{\cal{G}}_{jk}^{<}(t,t^\prime)
&&=i\langle A^{+}_{Tj}(\vec{k},t)A^{-}_{Tk}(-\vec{k},t^\prime)\rangle
\label{noneqgf2}\\\nonumber\\\nonumber
&&=\left[\frac{P_{jk}}{2}\frac{1}{2ik}\left\{
(1+n_k)V_{1k}^*(t)V_{1k}(t^\prime)
+n_kV_{1k}(t)V_{1k}^*(t^\prime)\right.\right.\\\nonumber\\\nonumber
&&\left.\left.
\hspace{0.5 in}+(1+n_k)V_{2k}^*(t)V_{2k}
(t^\prime)+n_kV_{2k}(t)V_{2k}^*(t^\prime)
\right\}\right.
\\\nonumber\\\nonumber
&&\left.
-\frac{f_{jk}}{2}\frac{1}{2k}\left\{(1+n_k)V_{1k}^*(t)V_{1k}(t^\prime)
+n_kV_{1k}(t)V_{1k}^*(t^\prime)\right.\right.\\\nonumber\\\nonumber
&&\left.\left.
\hspace{.5 in}-(1+n_k)V_{2k}^*(t)V_{2k}
(t^\prime)-n_kV_{2k}(t)V_{2k}^*(t^\prime)
\right\}\right].
\end{eqnarray}

Here, $\vec{A}^+_T$ and $\vec{A}^-_T$ represent the transverse components of
the gauge field 
 along the forward and backward time contours of the closed time path integral.

\newpage
\begin{figure}
\epsfxsize=6in
\epsfig{file=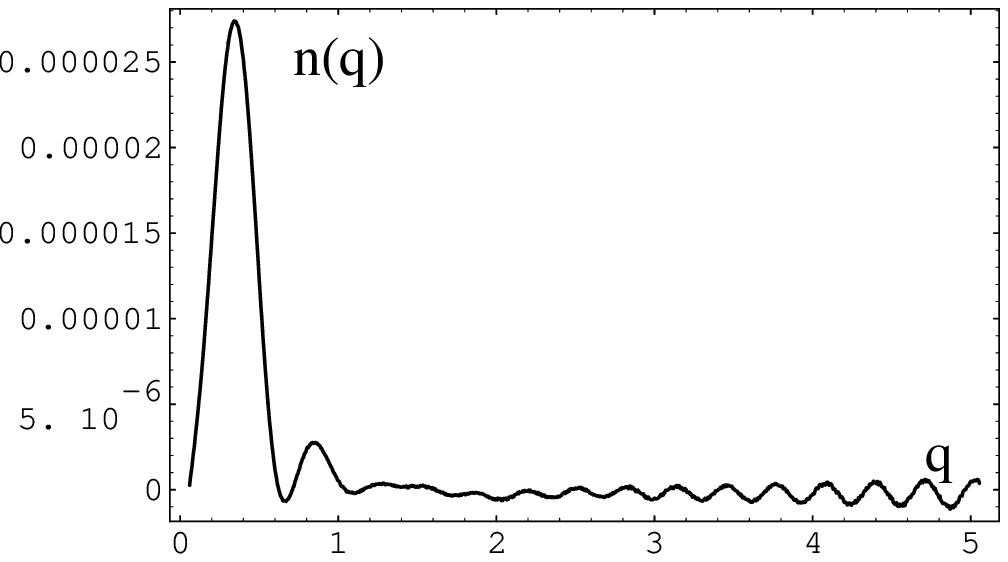}
\caption{$n(q,\tau)=(2\pi)^3|q|\frac{d(N(\tau)-N(0))/\Omega}{d^3q}$ (in units
of 
$\text{fm}^{-1}$), vs. $q$,
at time  $t=10$ fm/c, after subtracting the thermal background at
$T_i=220$ MeV, for  $\xi(0)=0.1$, $\dot{\xi}(0)=0$, 
$\phi(0)=\dot{\phi}(0)=0$, $T_i=220$ MeV.}
\epsfxsize=6 in 
\epsfig{file=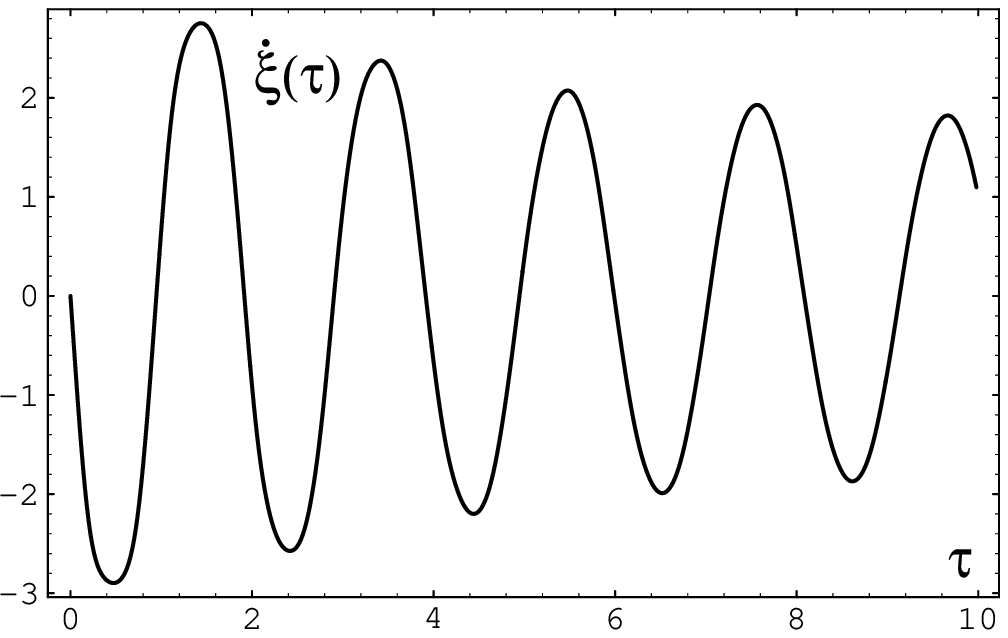} 
\caption{(a) $\dot{\xi}(\tau)$ vs. $\tau$ (in units
of fm/c) for
$\xi(0)=1.0$, $\dot{\xi}(0)=0$, 
$\phi(0)=\dot{\phi}(0)=0$, $T_i=220$ MeV.}
\vspace{2 in}
\addtocounter{figure}{-1}
\epsfxsize=6 in
\epsfig{file=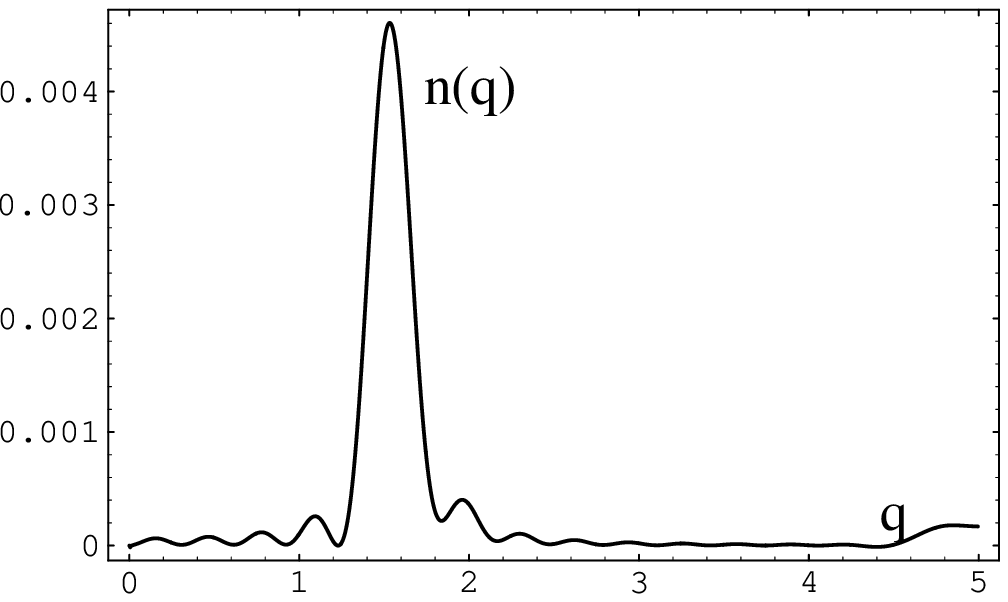}
\caption{(b) $n(q,\tau)=(2\pi)^3|q|\frac{d(N(\tau)-N(0))/\Omega}{d^3q}$ (in
units of 
$\text{fm}^{-1}$), vs. $q$,
at  time $t=10$ fm/c for the same initial conditions
as in Fig.(2.a).} 
\addtocounter{figure}{-1}
\epsfxsize=6 in
\epsfig{file=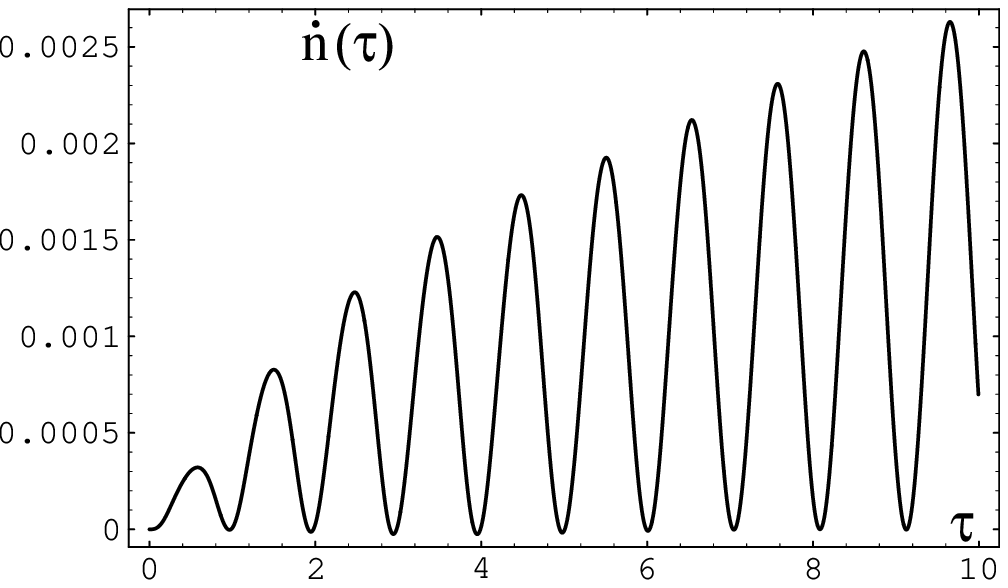}
\caption{(c) $\dot{n}(q,\tau)=(2\pi)^3|q|\frac{d\dot{N}(\tau)/ \Omega}{d^3q}$
(in 
units  of $\text{fm}^{-2}$), at the center of the
resonant band $qM_F=k=k_{res}=306$ MeV as a function of $\tau$ (in units of
fm/c),
for the same initial conditions as in Fig.(2.a).}
\epsfxsize=6 in
\epsfig{file=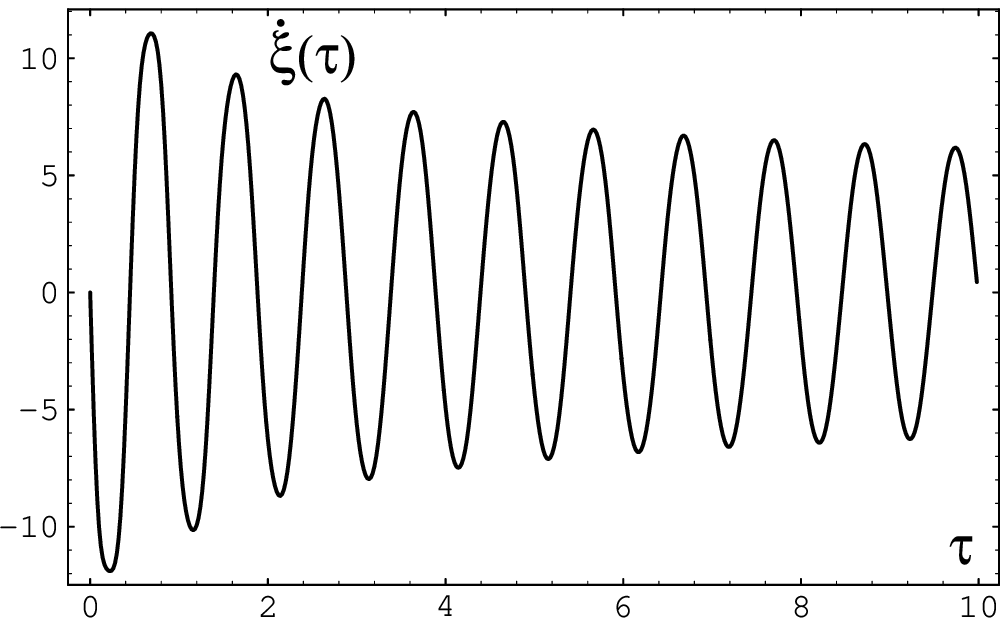}
\caption{(a) $\dot{\xi}(\tau)$ vs. $\tau$ (in units
of fm/c) for
$\xi(0)=2.0$, $\dot{\xi}(0)=0$, 
$\phi(0)=\dot{\phi}(0)=0$, $T_i=220$ MeV.}
\vspace{2 in}
\addtocounter{figure}{-1}
\epsfxsize=6 in
\epsfig{file=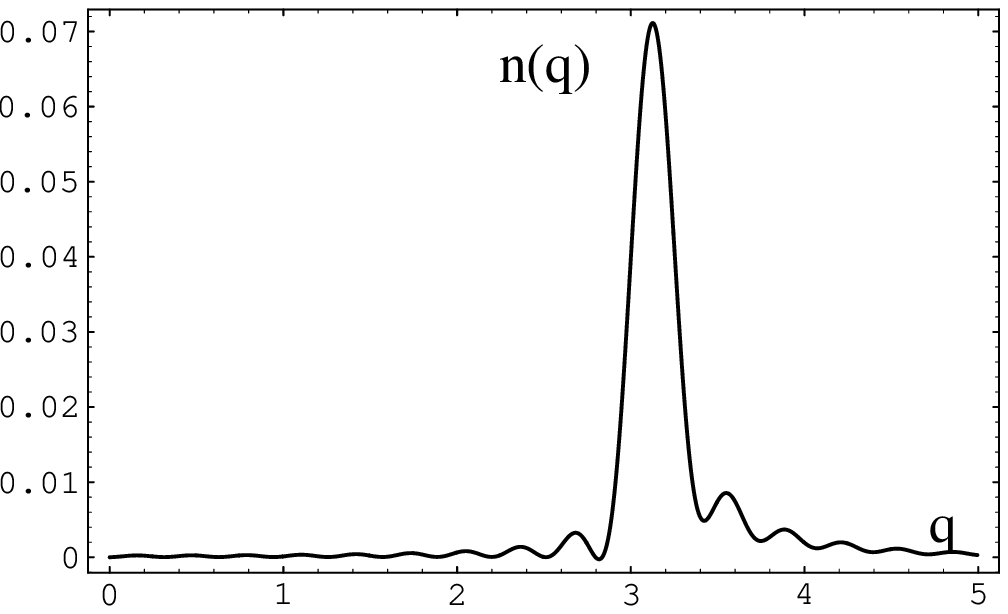} 
\caption{(b) $n(q,\tau)=(2\pi)^3|q|\frac{d(N(\tau)-N(0))/\Omega}{d^3q}$ (in
units of 
$\text{fm}^{-1}$), vs. $q$,
at time $t=10$ fm/c for the same initial conditions
as in Fig.(3.a).} 
\addtocounter{figure}{-1}
\epsfxsize=6 in
\epsfig{file=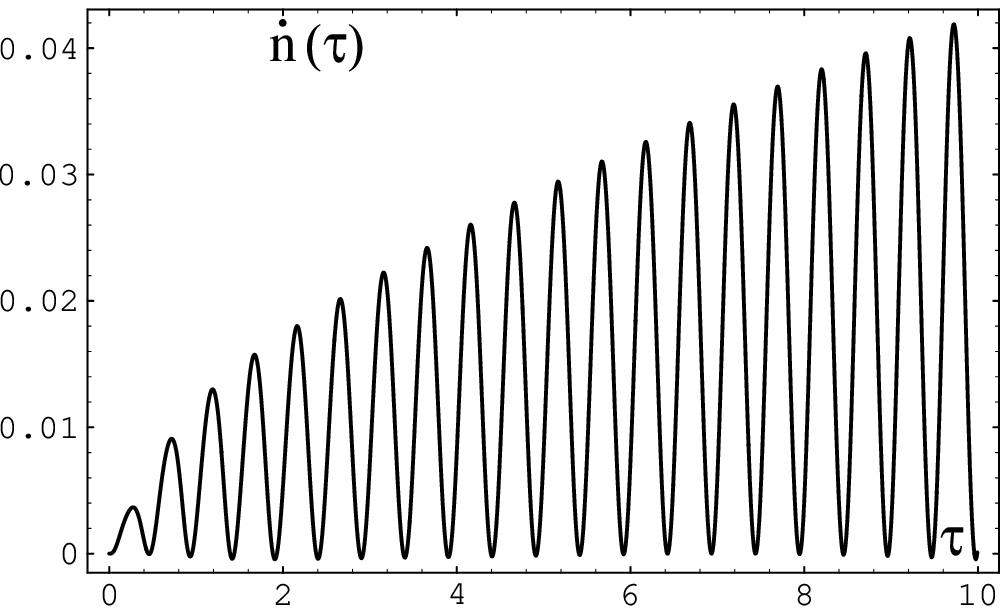} 
\caption{(c) $\dot{n}(q,\tau)=(2\pi)^3|q|\frac{d\dot{N}(\tau)/ \Omega}{d^3q}$ (in
units of  $\text{fm}^{-2}$), at the center of the
resonant band  $qM_F=k=k_{res}=625$ MeV as a function of $\tau$ (in units of
fm/c), 
for the same initial conditions as in Fig.(3.a).}
\epsfxsize=6 in
\epsfig{file=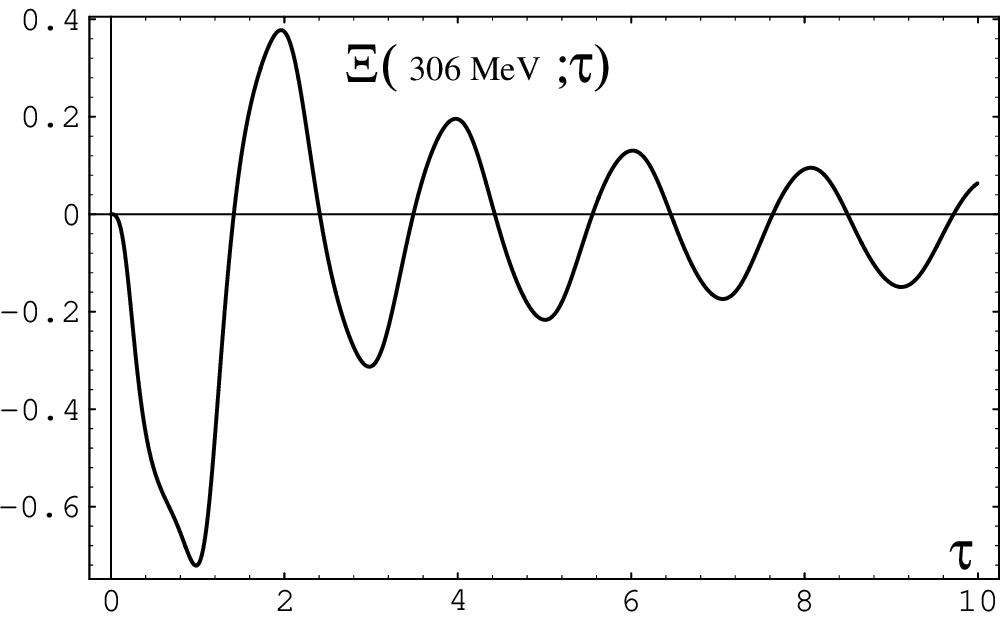}
\caption{(a) $\Xi(k=k_{res},\tau)=(n_+-n_-)/(n_++n_-)$ vs. $\tau$ (in units of
fm/c), for  $k=306$ $\text{MeV}$ and for the same initial conditions as
in Fig.(2.a).} 
\addtocounter{figure}{-1}
\epsfxsize=6 in
\epsfig{file=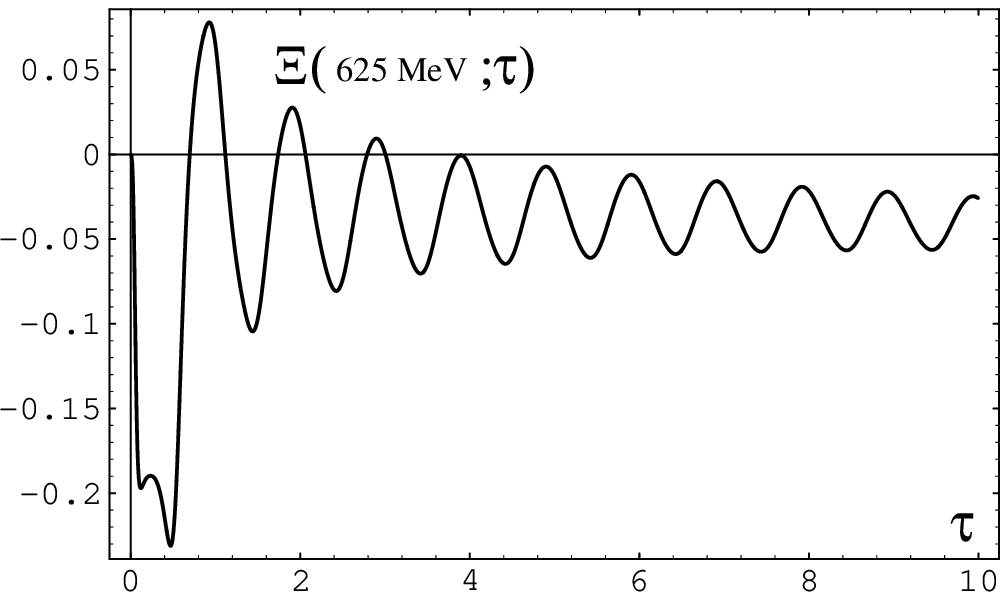}
\caption{(b) $\Xi(k=k_{res},\tau)=(n_+-n_-)/(n_++n_-)$ vs. $\tau$ (in units of
fm/c), for $k=625$ $\text{MeV}$ and for the same initial conditions as
in Fig.(3.a).}
\epsfxsize=6 in
\epsfig{file=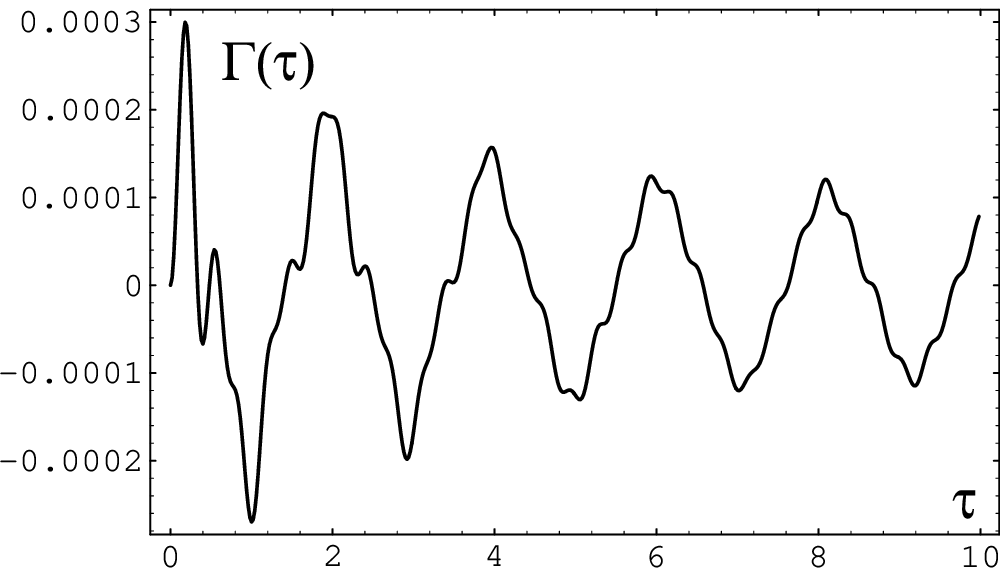} 
\caption{(a) $\Gamma(\tau)=\alpha M_F\langle F\tilde{F}\rangle/8\pi f_{\pi}$
(in units
of $M^4_F$), vs. $\tau$ (in units of fm/c) for the
initial conditions of Fig.(2.a).}
\addtocounter{figure}{-1}
\epsfxsize=6 in
\epsfig{file=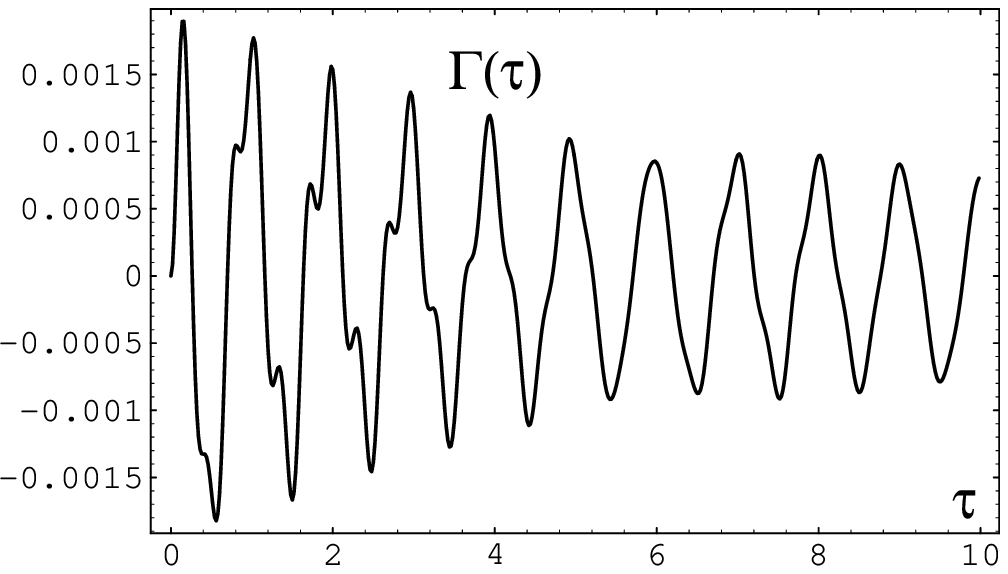}
\caption{(b) $\Gamma(\tau)=\alpha M_F\langle F\tilde{F}\rangle/8\pi f_{\pi}$
(in units 
of $M^4_F$), vs. $\tau$ (in units of fm/c) for the
initial conditions of Fig.(3.a).} 
\epsfxsize=6 in
\epsfig{file=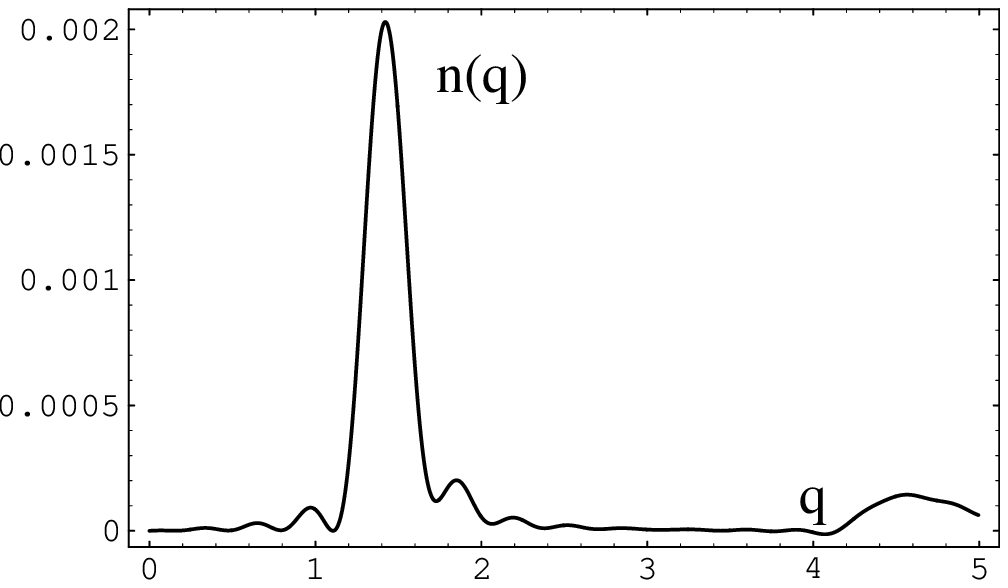}
\caption{(a) $n(q,\tau)=(2\pi)^3|q|\frac{d(N(t)-N(0))/\Omega}{d^3q}$ (in units
of $\text{fm}^{-1}$), vs. $q$ at time $t=10$ fm/c 
 for $\xi(0)=1.0$, $\dot{\xi}(0)=0$, 
$\phi(0)=\dot{\phi}(0)=0$, $T_i=0$.}
\addtocounter{figure}{-1}
\epsfxsize=6 in
\epsfig{file=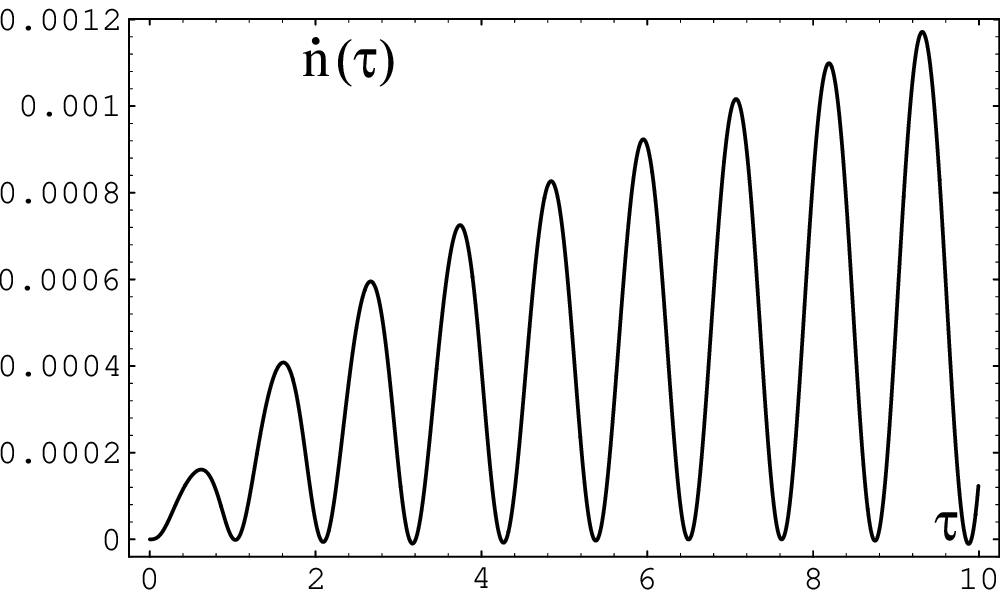}
\caption{(b) $\dot{n}(q,\tau)=(2\pi)^3|q|\frac{d\dot{N}(t)/ \Omega}{d^3q}$  (in
units of 
$\text{fm}^{-2}$), at the center of the resonant band 
$qM_F=k=k_{res}=284$ MeV
as a function of $\tau$ (in units of fm/c) for the same initial
conditions as in Fig.(6.a).}
\addtocounter{figure}{-1}
\epsfxsize=6 in
\epsfig{file=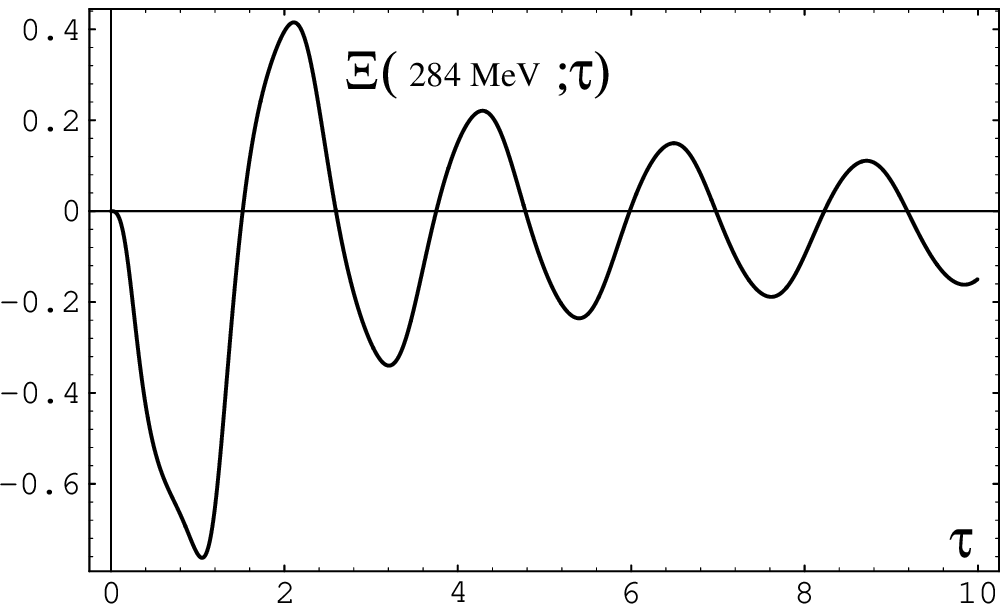}
\caption{(c) $\Xi(k=k_{res},\tau)=(n_+-n_-)/(n_++n_-)$ vs. $\tau$ 
(in units of fm/c), for $k=284$ MeV and for the same initial conditions as in
Fig.(6.a).} 
\end{figure}
\end{document}